\documentclass[11pt,a4paper]{article}
\pdfoutput=1

\usepackage{dcolumn}
\usepackage{bm}
\usepackage{amsthm}
\usepackage{setspace}
\usepackage[normalem]{ulem}
\usepackage{relsize}
\usepackage{cancel}
\usepackage{verbatim}
\usepackage{enumitem}
\usepackage{mathtools}
\usepackage{empheq}
\usepackage{dsfont}
\usepackage{amsmath}
\usepackage{slashed}

\usepackage{graphicx}
\usepackage{caption}
\usepackage{subcaption}

\usepackage{jheppub}

\usepackage{amsthm}

\newcommand{\del}{\partial}

\newcommand{\tr}{\operatorname{Tr}}

\newcolumntype{I}[1]{>{\centering\arraybackslash$}m{#1}<{$}}

\title{On Supersymmetric Lifshitz Field Theories}

\author{Shira Chapman,}
\author{Yaron Oz,}
\author{Avia Raviv-Moshe}
\affiliation{Raymond and Beverly Sackler School of Physics and Astronomy, Tel-Aviv University, 55 Haim Levanon street, Tel-Aviv, 69978, Israel}
\emailAdd{shirator@post.tau.ac.il}
\emailAdd{yaronoz@post.tau.ac.il}
\emailAdd{aviaravi@mail.tau.ac.il}

\abstract{
We consider field theories that exhibit a supersymmetric Lifshitz scaling with  two  real supercharges. The theories
 can be formulated  in the language of stochastic quantization.
We construct the free field supersymmetry algebra with rotation singlet fermions for an even dynamical exponent $z=2k$ in an arbitrary dimension.
We analyze the classical  and quantum $z=2$ supersymmetric interactions  in $2+1$ and $3+1$ spacetime dimensions and 
reveal a supersymmetry preserving quantum diagrammatic cancellation. 
Stochastic quantization indicates that Lifshitz scale invariance is broken in the $(3+1)$-dimensional quantum theory.
}


\begin{document}
\maketitle
\flushbottom

\section{Introduction}

Lifshitz scaling is a symmetry under which time and space scale differently:
\begin{equation}\label{Intro:LifshitzScaling}
t \rightarrow \lambda^{-z} t \qquad  x^i\rightarrow \lambda^{-1} x^i \qquad i=1,\dots,d \, ,
\end{equation}
where $d$ is the number of space dimensions and $z$ is the dynamical critical exponent, which measures the anisotropy between space and time. 
While there is no sign for Lifshitz scaling at high energy, it is a property of certain low energy systems of condensed matter, that exhibit quantum criticality
(see e.g. \cite{Sachdev:2011cup}). 

The generators of Lifshitz symmetry in $d+1$ spacetime dimensions are time translation $P_0$, space translations $P_i$, the scale transformation 
$D$ and space rotations $M_{ij}$. The Lifshitz algebra has the following commutation relation structure: 
\begin{align}
\begin{split}
&[D, P_i] = i  P_i  , \quad [D,P_0] = i  z P_0 , \quad   [M_{ij}, M_{kl}] = i \delta_{kj} M_{il}+ \dots  ,
\\
&  
[M_{ij}, P_k] = - i \delta_{ki}P_{j} +\dots ,    ~\quad \qquad  [M_{ij}, P_0] = 0  .
\end{split}
\end{align}
Note, that the Lifshitz algebra has no Casimir operators that are polynomial in the generators, and thus has no nontrivial irreducible representations. 

In this work we will study supersymmetric extensions of the Lifshitz symmetry, in which the dynamical exponent is even. 
Supersymmetry involves both  Grassmann odd and Grassmann even generators. With boost invariance,
they are associated by the spin-statistics theorem
to fermions and bosons, respectively. Since boosts are not part of the Lifshitz algebra, this is no longer true. 
We will denote by $Q$ the Grassmann odd generators of the Lifshitz superalgebra. They are  characterized by their
representation under the $d$-dimensional rotation group, and we will consider the singlet representation.

In the relativistic $z=1$ case, one has $\left\{  Q , Q \right\} \sim \slashed{P}$. Here, however, the dimensions of $P_0$ and $P_i$ are
different and this can no longer be the case.
Using a free field realization with singlet fermions,  we find that the supersymmetric algebra has the structure:
\begin{equation}
\left\{  Q , Q \right\} \sim H, \quad [M_{ij} , Q_\alpha ]=0 , \quad [ D , Q ] = i  \frac{z}{2} Q \ .
\end{equation}

Supersymmetry possesses attractive features in the context of high energy particle physics. Such are the cancellation of quadratic divergences
in the radiative corrections to the Higgs boson mass, the unification of the standard model gauge couplings, and having a natural candidate for dark matter (see e.g. \cite{Martin:1997ns} and references therein).
By now, supersymmetry is the most extensively studied generalization of the standard model. In the context of low energy condensed matter systems, models satisfying the detailed balance condition admit a natural supersymmetric generalization \cite{Parisi:1982ud,Dijkgraaf:2009gr,Orlando:2009az,Horava:2008ih}. These often arise in the context of stochastic quantization rendering supersymmetry natural in models with a stochastic noisy background. 
Yet, both at low energy and high energy, supersymmetry has not been found experimentally.
Previous studies of supersymmetry in Lifshitz field theories are \cite{Orlando:2009az,Gomes:2014tua,Xue:2010ih}.

We will consider in detail supersymmetric parity and time reversal invariant interactions for $z=2$ in $2+1$ and $3+1$ spacetime dimensions. These interactions can be constructed  using stochastic quantization with a detailed balance condition.
We will 
study the quantum corrections at first and partially second order in perturbation theory. Supersymmetry is preserved
quantum mechanically and we
reveal a cancellation mechanism lowering the degree of the diagrammatic divergence.

The paper is organized as follows. 
In section \S \ref{sec:FreeLifshitzSUSY} we construct  a free Lifshitz Wess-Zumino model  and the supersymmetry
algebra for even z and singlet fermions. In section \S \ref{sec:Interactions} we 
outline the supersymmetric invariant interactions for $z=2$ in $2+1$ and $3+1$ spacetime 
dimensions, and in section \S \ref{sec:QuantumCorrections} we analyze the quantum corrections. We conclude with a summary and discussion in \S \ref{sec:Sum}. A list
 of notations and conventions is given in appendix \S \ref{app:conventions}.

\section{Free Supersymmetric Lifshitz Field Theories}\label{sec:FreeLifshitzSUSY}
In this section we will construct free supersymmetric Lifshitz field theories and the corresponding supersymmetry algebra.

\subsection{Lifshitz Wess-Zumino Model}\label{subsec:free:wz}

We consider the following Lifshitz action for $z = 2k$ and singlet fermions (the summary of conventions can be found in appendix \ref{app:conventions}):
\begin{equation}\label{eq:actionfree}
S = \int dt{d^d}x\left[ (\del_t\phi)^2 - g^2 \left( \nabla ^{2k}\phi  \right)^2+ \bar \psi \gamma^0 \del_t \psi  + g \bar \psi \nabla ^{2k} \psi  + F^2 \right] \ .
\end{equation}
$\phi$ is a real boson,  $\psi_{\alpha}$ is a two component (one on-shell degree of freedom) real fermion field  $\bar \psi = \psi^T \sigma^2$ where $\sigma^2$ is the second Pauli matrix, and
$F$ is an auxiliary bosonic field. $g$ is a dimensionless parameter measuring the relative strength of the kinetic terms. It
is unconstrained in the absence of boost invariance. 

The scaling dimensions of the fields are given by
$[\phi]=(d-z)/2$, $[\psi] = \frac{d}{2}$ and $[F]=(d+z)/2$.
In a supersymmetric model one expects the bosons and fermions to have the same dispersion relation, hence the relation between the coefficients of the spatial parts of their kinetic terms.
Note, that singlet fermions seem to be incompatible with odd $z$, since it is not clear how to write a local and rotationally invariant Lagrangian.

The action \eqref{eq:actionfree} is invariant under the following supersymmetry transformation:
\begin{equation}
\label{Free_SUS:SusyTrans}
\begin{split}
&\delta \phi  = \bar \epsilon \psi ,
\\
&\delta \psi  = \left( { - {\gamma ^0}{\partial _t} + g{\nabla ^{2k}}} \right)\phi \epsilon  + F\epsilon, 
\\
& \delta \bar \psi  = \bar \epsilon \left( {{\gamma ^0}{\partial _t} + g{\nabla ^{2k}}} \right)\phi  + \bar \epsilon F,
\\
& \delta F = \bar \epsilon \left( {-{\gamma ^0}{\partial _t} - g{\nabla ^{2k}}} \right)\psi \ ,
\end{split}
\end{equation}
with the commutation of two supersymmetry transformations given by:
\begin{equation}
\left(\delta_{\epsilon_1} \delta_{\epsilon_2} - \delta_{\epsilon_2}\delta_{\epsilon_1} \right)X = - 2{{\bar \epsilon }_2}{\gamma ^0}{\epsilon _1}{\partial _t}X,
\end{equation}
and $X\equiv\{\phi,\psi,F \}$.

The Noether supercurrents associated with the supersymmetry transformation \eqref{Free_SUS:SusyTrans} read:
\begin{equation}
\begin{split}\label{eq:SupCur}
&J^0 = 2 \left( {\bar \psi {\partial _t}\phi  + g \bar \psi {\gamma ^0}{\nabla ^{2k}}\phi } \right),
\\
&  J^i = - 2 g \sum\limits_{n=0}^{2k-1} (-1)^n  
\left(\del_i^n \bar \psi \gamma^0 \del_i^{2k-n-1} \del_t \phi -g \del_i^n \bar \psi \del_i^{4k-n-1} \phi  \right) \ ,
\end{split}
\end{equation}
and using the canonical commutation relations between the fields:
\begin{equation}
\label{eq:ComAnti}
\left\{ {{\psi _\alpha }\left( \vec x \right),{{\bar \psi }_{\beta }}\left(\vec y \right)} \right\} = -\frac{i}{2}\left(\gamma ^0\right)_{\alpha \beta }{\delta ^2}\left( {\vec x - \vec y} \right),~~~
\left[ {\phi \left(\vec x \right),\dot \phi \left( \vec y \right)} \right] = \frac{i}{2}{\delta ^2}\left( {\vec x -\vec  y} \right)  \ , 
\end{equation}
one  gets the supersymmetry algebra:
\begin{equation}
\label{eq:Alg1}
\left\{  Q_\alpha , Q_\beta  \right\} = 2\gamma ^0_{\alpha \beta } H \ ,
\end{equation}
where the supercharges are given by:
\begin{equation}
\label{eq:SupCharge}
Q= 2 \int {{d^d}x}  \left( {\bar \psi {\partial _t}\phi  + g \bar \psi {\gamma ^0}{\nabla ^{2k}}\phi } \right) \ ,
\end{equation}
and the Hamiltonian reads:
\begin{equation}
\label{eq:Ham}
H = \int d^d x 
\left( \left(\del_t \phi \right)^2 + g^2 \left( \nabla^{2k}\phi  \right)^2 - g \bar\psi \nabla^{2k} \psi \right) .
\end{equation}
In two space dimensions the action (\ref{eq:actionfree}) is invariant under the supersymmetry transformations (\ref{Free_SUS:SusyTrans})
even when the fermions 
are not singlets. The reason being that one can use the additional $SO(2)$ fermion particle number symmetry to relate the different covers of the spatial
rotation group.
It is also  straightforward to check, at least in two space dimensions,  that the action (\ref{eq:actionfree}) and the supersymmetry transformations (\ref{Free_SUS:SusyTrans}) compose the most general supersymmetric structure with singlet fermions and $SO(2)$ particle number symmetry.

In the derivation of the commutation relation \eqref{eq:Alg1} it is useful to use the rotated complex spinors:
\begin{equation}\label{complex_fermions_def}
\lambda_1 \equiv \frac{\psi_1-i\psi_2}{2},~~~~~~
\lambda_2 \equiv \lambda_1^* = \frac{\psi_1+i\psi_2}{2} \ ,
\end{equation}
that satisfy the canonical anticommutation relations:
\begin{equation}
\left\{ \lambda_1(\vec x,t) , \lambda_2(\vec y,t) \right\} = \frac{1}{4}\delta^2(\vec x- \vec y) \ .
\end{equation}
The fermionic Lagrangian reads:
\begin{equation}
\mathcal{L}_{f} = 4i\lambda_1 \del_t \lambda_2 +4g \lambda_1 \nabla^{2k} \lambda_2 \ ,
\end{equation}
and the fermion particle number symmetry is:
\begin{equation}\label{U_1}
\lambda_1 \rightarrow  e^{i\theta} \lambda_1, \qquad \lambda_2 \rightarrow e^{-i \theta} \lambda_2 \ .
\end{equation}

The free supersymmetric action for the singlet fermions and the scalar can we written using the 
stochastic quantization approach (see e.g. \cite{Damgaard:1987rr}), where:
\begin{equation}
\label{DetailedBalance3}
 S\left[ {\phi ,\psi ,\bar \psi } \right] = \int {{d^d}xdt} \left( {{{\dot \phi }^2} - {{\left( {\frac{{\delta W}}{{\delta \phi }}} \right)}^2} + \bar \psi \left( \gamma^0 {\frac{d}{{dt}} - \frac{{{\delta ^2}W}}{{\delta {\phi ^2}}}} \right)\psi } \right) \ ,
\end{equation}
and
\begin{equation}
\label{DetailedBalance4}
W\left( \phi  \right) = -\frac{g}{2} \int d^d x \left( \phi \nabla^{2k} \phi \right)  .
\end{equation} 
This will prove useful when constructing the supersymmetric invariant interactions.

\section{$z=2$ Supersymmetric Interactions}\label{sec:Interactions}

In this section we construct  $z=2$ supersymmetric parity and  time reversal invariant
interaction terms in $2+1$ and $3+1$ spacetime dimensions. We assume the existence of an $SO(2)$ particle number symmetry  \eqref{U_1}.
Consider the parity transformation on one coordinate ${x_1} \to  - {x_1}$ of the vector $\vec x =(x_1,\dots,x_d)$. We define the action on the fermionic field, such that the free supersymmetric
Lagrangian of the previous section is invariant.
It reads
\begin{equation}
\label{ParityFermion}
P_1 \psi(t,x_1,x_2, \dots,x_d) P_1 = \psi (t,-x_1,x_2, \dots ,x_d) \ .
\end{equation}
Time reversal takes $i \rightarrow -i$, $\del_t \to - \del_t$ and we similarly define 
\begin{equation}
\label{TimeRevFermion}
T\psi (t,\vec x) T = - {\gamma _2} \psi (-t,\vec x) \ ,
\end{equation}
or alternatively with $\gamma_1$. In terms of the independent components we have:
\begin{equation}
\begin{split}
T\psi_1 (t,\vec x) T = \psi_1 (-t,\vec x),
\qquad
T\psi_2 (t,\vec x) T = -\psi_2 (-t,\vec x),
\\
T\lambda_1 (t,\vec x) T = \lambda_1 (-t,\vec x),
\qquad
T\lambda_2 (t,\vec x) T = \lambda_2 (-t,\vec x).
\end{split}
\end{equation}
The most general local power counting (in weighted Lifshitz units) renormalizable\footnote{For a study of renormalizability of Lifshitz theories see e.g. \cite{Anselmi:2007ri}.}  interactions invariant under parity, time reversal and SO(2) particle number symmetries take the form:
\begin{equation}
\label{Interactions:IntGene}
\mathcal{L}_{\mathop{\rm int}} = \bar \psi \psi W + F\bar \psi \psi W_1 + F W_2 + F^2 W_3  + W_4 + \bar \psi \gamma ^0 \partial _t \psi W_5 + \bar \psi \nabla ^2\psi W_6   ,
\end{equation}
where the $W$, $W_i$ denote real polynomial functions of the boson $\phi$ and its derivatives. 
We will consider two sets of interactions, marginal and relevant.
We find that unlike the relativistic case here we only have derivative interactions.

\subsection{Marginal Interactions in $2+1$ Dimensions}

In this subsection we study interactions whose parameters are dimensionless in weighted Lifshitz units which are invariant under supersymmetry.\footnote{The parity and time reversal odd interaction  $\epsilon^{ij} \bar\psi \gamma^0 \nabla_i \psi \nabla_j \phi$ of \cite{Fitzpatrick:2012ww} is not invariant under supersymmetry due to the $\epsilon^{ij} \bar\psi \gamma^0 \nabla_i \psi \bar \epsilon \nabla_j \psi$ contribution to the supersymmetric variation which cannot be cancelled. Hence in the presence of supersymmetry this term cannot serve as a mechanism for creating anyons.} 
The most general parity and time reversal invariant marginal interaction is a linear combinations of the following two sets of interactions --
marginal interactions without time derivatives:
\begin{equation}\label{Interactions:Set1}
\begin{split}
\mathcal{L}_{{\rm{int}}}^a = \sum_{n = 1}^\infty  
a_n \phi^{n - 1} & \left[ - g F  \left( { 2\phi {\nabla ^2}\phi  + n{\nabla _i}\phi {\nabla ^i}\phi } \right)   \right.  \\
& ~~\left.  + \frac{n}{2}  g  \bar \psi \psi \nabla ^2 \phi  + g \bar \psi {\nabla ^2}\psi \phi - 2 g^2 \phi  \left( \nabla ^2 \phi \right)^2 - n g^2 {\nabla ^2}\phi {\nabla _i}\phi {\nabla ^i}\phi  \right] \ ,
\end{split}
\end{equation}
and marginal interactions containing time derivatives of the fields:
\begin{equation} \label{Interactions:Set2}
\begin{split}
\mathcal{L}_{{\rm{int}}}^b =  \sum_{n = 1}^\infty   b_n \phi^{n-1}
 & \left[ {{F^2}\phi +F\left(2 g \phi {\nabla ^2}\phi - \frac{n}{2} \bar \psi \psi  \right) } \right.  \\ 
&~~ \left.  - \frac{n}{2} g\bar \psi \psi {\nabla ^2}\phi  + \bar \psi {\gamma ^0}{\partial _t}\psi \, \phi + g^2  \phi \left( \nabla ^2 \phi \right)^2  + \phi (\del_t \phi)^2 \right],
\end{split}
\end{equation}
where $a_n$ and $b_n$ are coefficients that do not depend on the fields. This is derived by writing
the most general form of the functions $W$, $W_i$  and requiring invariance under the supersymmetry transformations. 

One could at this point integrate out the auxiliary field using its equation of motion.
Note that in the case of $\mathcal{L}_\text{int}^b$ this solution turns out to be non-local in the fields. We will proceed with the first set of interactions only and leave the second set (with time derivatives) for future study.
Solving for the auxiliary field we get
\begin{equation}
\label{eq:a2}
\begin{split}
\mathcal L^a = & {\left( {{\partial _t}\phi } \right)^2} - g^2 {\left( {{\nabla ^2}\phi } \right)^2} + \bar \psi {\gamma ^0}{\partial _t}\psi  + g \bar \psi {\nabla ^2}\psi
\\
 &~~~~
- 
 \left( \sum\limits_{n = 1}^\infty  
a_n g \left( { \phi {\nabla ^2}\phi  + \frac{n}{2}{\nabla _i}\phi {\nabla ^i}\phi } \right) \phi^{n - 1}  \right)^2  \\ 
 &~~~~  + \sum\limits_{n = 1}^\infty  
a_n  g \phi^{n - 1}  \left[\frac{n}{2} { \bar \psi \psi \nabla ^2}\phi  + \bar \psi {\nabla ^2}\psi \phi - 2 g \phi  \left( \nabla ^2 \phi \right)^2  - n g {\nabla ^2}\phi {\nabla _i}\phi {\nabla ^i}\phi 
\right] .
 \end{split}
\end{equation}

\subsection{Relevant Deformations in $2+1$ Dimensions}
In this section we study supersymmetric 
deformations of Lifshitz theories which do not include fermion derivatives. This means that we will look at the interaction \eqref{Interactions:IntGene} with $W_5$ and $W_6$ set to zero. Under this condition we find that the most general supersymmetric interaction takes the form:
\begin{equation}
\label{eq:Set3}
\mathcal L_{{\rm{int}}}^c = \sum\limits_{n = 1}^\infty  {{c_n}{\phi ^n}F - \frac{1}{2}{c_n}n{\phi ^{n - 1}}\bar \psi \psi  - {c_n}n \, g {\phi ^{n - 1}}{\nabla _i}\phi {\nabla ^i}\phi } .
\end{equation}
These interactions have coupling constants with positive Lifshitz scaling dimension $\left[ c_n \right] = 2 $. Choosing for example $n=2$ one gets a Yukawa-like interaction. This is derived as follows. Requiring cancellation of $F^2$ terms and terms which contain one auxiliary field, one fermion and one time derivative after the supersymmetry variation we obtain $W_1=W_3=0$. Cancellation of three fermion terms requires $W$ to be a polynomial in the bosonic field. $W_2$ and $W_4$ can then be found by making the most general ansatz with arbitrary coefficients and imposing invariance under supersymmetry up to total derivatives. Solving for the auxiliary field and rearranging we get
\begin{equation}
\begin{split}
\mathcal L^c = & {\left( {{\partial _t}\phi } \right)^2} - g^2 {\left( {{\nabla ^2}\phi } \right)^2} + \bar \psi {\gamma ^0}{\partial _t}\psi  + g \bar \psi {\nabla ^2}\psi  \\ 
 &- \frac{1}{4}{\left( {\sum\limits_{n = 1}^\infty  {{c_n}{\phi ^n}} } \right)^2} - \frac{1}{2}\sum\limits_{n = 1}^\infty  {{c_n}n{\phi ^{n - 1}}} \bar \psi \psi  + \sum\limits_{n = 1}^\infty  c_n  g  \phi^n {\nabla^2}\phi.
 \end{split}
\end{equation}
Note that the $n=1$ terms have a quadratic field dependence. We will therefore often denote $c_1\equiv 2 m^2$ and write the Lagrangian as:\footnote{$m$ is not a mass in the standard non-relativistic sense (this role is played by $g$), but rather in the sense that it serves as an IR regulator.}
\begin{equation}
\begin{split}\label{eq:c2}
\mathcal L^c = &  (\partial_t \phi)^2 - g^2 \left(\nabla ^2\phi\right)^2 + 2m^2 g \phi {\nabla ^2}\phi  - m^4{\phi ^2} + \bar \psi {\gamma ^0}{\partial _t}\psi  + g \bar \psi {\nabla ^2}\psi  - m^2\bar \psi \psi  
\\
&
- \sum\limits_{n,m = 2}^\infty \frac{c_n c_m}{4} \phi ^{n+m} 
- m^2 \sum\limits_{n = 2}^\infty  c_n \phi ^{n + 1}
- \sum\limits_{n = 2}^\infty  \frac{n}{2}  c_n \phi ^{n - 1} \bar \psi \psi 
+ \sum\limits_{n = 2}^\infty  c_n  g  \phi^n {\nabla^2}\phi.
 \end{split}
\end{equation}

\subsection{Interactions in 3+1 Dimensions}
We keep the $(2+1)$-dimensional notations of the previous subsections (see also appendix ~\ref{app:conventions}). In this model the fermion field transforms as a singlet under spatial rotations.
The $(3+1)$-dimensional fields have scaling dimensions of $\left[ \phi  \right] = \frac{1}{2}$, $[\psi] = \frac{3}{2}$ and $[F] = \frac{5}{2}$ respectively. Therefore, in order for the Lagrangian to have couplings of non-negative (weighted) dimensions, the interactions must consist of a finite  series in powers of $\phi$. The most general action in $3+1$ dimensions which is invariant under the supersymmetry transformation is captured by \eqref{Interactions:IntGene} with $W_1=W_3=W_5=W_6=0$.

 Imposing invariance under the supersymmetry transformation \eqref{Free_SUS:SusyTrans}:
\begin{equation}
\begin{split}\label{4dactionGen}
\mathcal{L}^{3 + 1} = (\partial_t \phi)^2 - g^2 \left(\nabla ^2\phi\right)^2 + 2m^2 g \phi {\nabla ^2}\phi  - m^4{\phi ^2} + \bar \psi {\gamma ^0}{\partial _t}\psi  + g \bar \psi {\nabla ^2}\psi  - m^2\bar \psi \psi  \\
 - m^2\sum\limits_{n = 2}^5 {c_n} {\phi ^{n + 1}} - \sum\limits_{n,m = 2}^5 {\frac{{c_n c_m}}{4}} {\phi ^{n + m}} + \sum\limits_{n = 2}^5 {c_n} {g}{\phi ^n}{\nabla ^2}\phi  - \sum\limits_{n = 2}^5 {\frac{{n }}{2}} c_n {\phi ^{n - 1}}\bar \psi \psi .
\end{split}
\end{equation}
This is the same as the $2+1$ dimensional relevant interactions with the infinite series cutoff at $n=5$. The couplings have weighted scaling dimensions of $[c_n]=\frac{5-n}{2}$ and $c_5$ represents a marginal interaction.

\subsection{The Detailed Balance Condition}\label{subsec:detbal}

The supersymmetric interactions can be built using the stochastic quantization approach and the detailed balance condition. A partial case containing only the first two terms in the infinite sum of the relevant interactions considered here was studied in \cite{Orlando:2009az}. By defining:
\begin{equation}
\label{DetailedBalance1}
W\left( \phi  \right) = \frac{1}{2} \int d^2x\left( g\left( \nabla _i \phi  \right)^2 - \sum\limits_{n=1}^{\infty}g a_n\frac{\phi ^{n + 1} \nabla ^2 \phi }{n + 1} + \sum\limits_{n=1}^{\infty} c_n \frac{\phi ^{n + 1}}{n + 1}  \right) \ ,
\end{equation} 
the most general action in $2+1$ dimensions which is invariant under our supersymmetry transformation can be calculated by
using (\ref{DetailedBalance3}).
One can also use this framework  to preform the quantum calculations in the $(1+1)$-dimensional  $z=1$ theory of a real bosonic field and check that the diagrammatical divergence structure agrees with our results (the detailed discussion is in  section ~\ref{subsec:Comp_Detailed_Balance}). Note, however, that the relative normalization of the kinetic terms cannot be obtained in this way \cite{Horava:2008ih}. Quantum mechanically we cannot separately fix the  renormalization of $\phi$ and $g$.
The same analysis can be performed for the $(3+1)$-dimensional scale invariant model by setting $a_n=0$ for all $n$, and $c_n=0$ for all $n>5$.

\section{Quantum Corrections}\label{sec:QuantumCorrections}

In this section we study the quantum corrections to the marginal and relevant interactions of the previous section (with coefficients $a_n$ and $c_n$ respectively). We find that in both cases there is a cancellation mechanism which reduces the naive degree of divergence at first order in perturbation theory. For the relevant interactions we also show this at second order in perturbation theory.
We compute the relations between the renormalized coupling constants at first order in perturbation theory and find that to this order the relations imposed by supersymmetry are not modified
and supersymmetry is not broken.
As we will see the relevant set of interactions cannot introduce corrections to the relative normalization of the kinetic terms $g$ and the marginal couplings $a_n$. This is expected on dimensional arguments. 

\subsection{Quantum Corrections to the Relevant Interactions in 2+1 Dimensions}\label{subsec:Rel_quant}

\begin{figure}
        \centering
        \begin{subfigure}[b]{0.2\textwidth}
                \includegraphics[width=\textwidth]{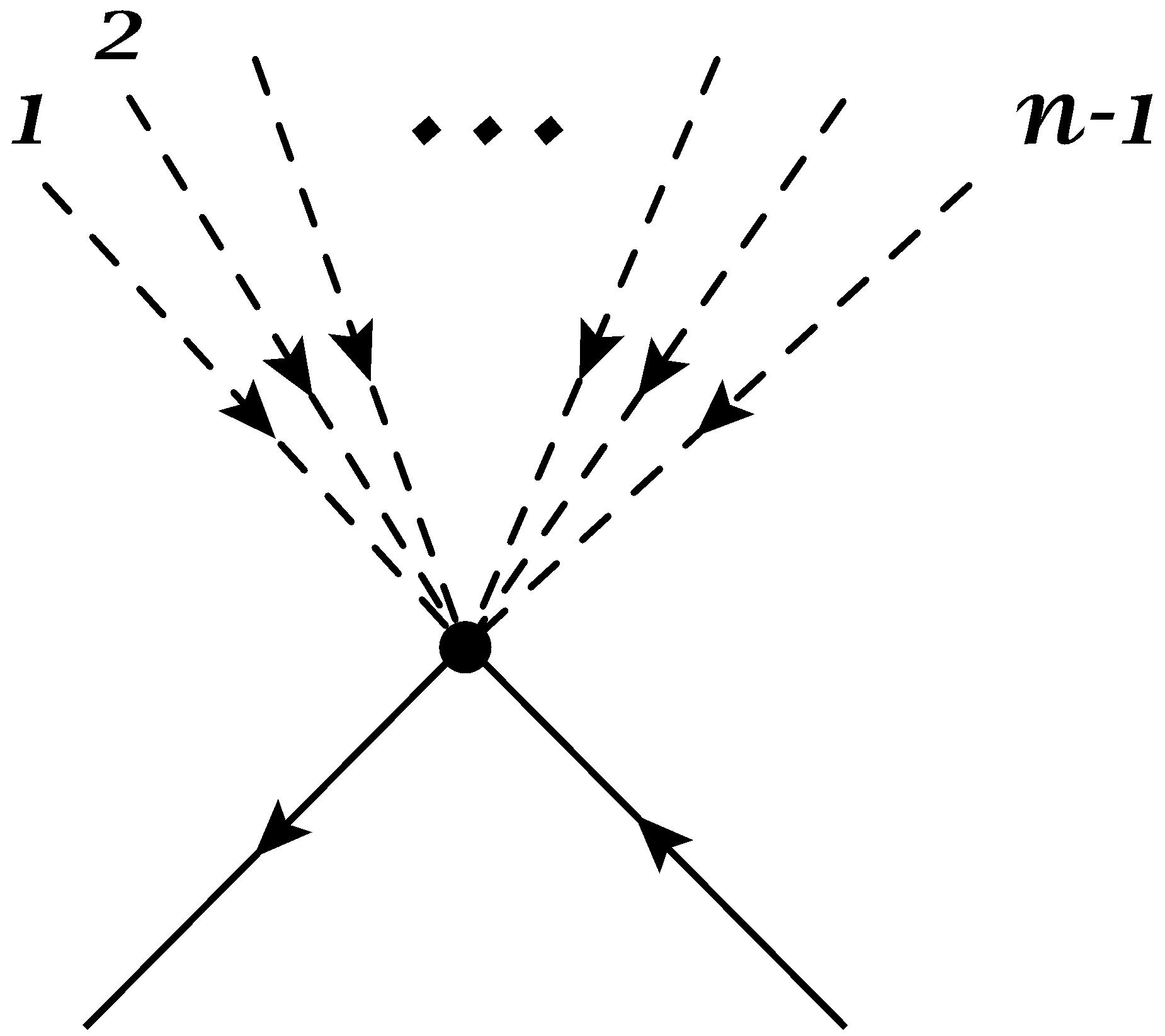}
                \caption{$\displaystyle - \frac{i}{2} n c_n $}
                \label{fig1:frb}
        \end{subfigure}
~\quad
        \begin{subfigure}[b]{0.2\textwidth}
                \includegraphics[width=\textwidth]{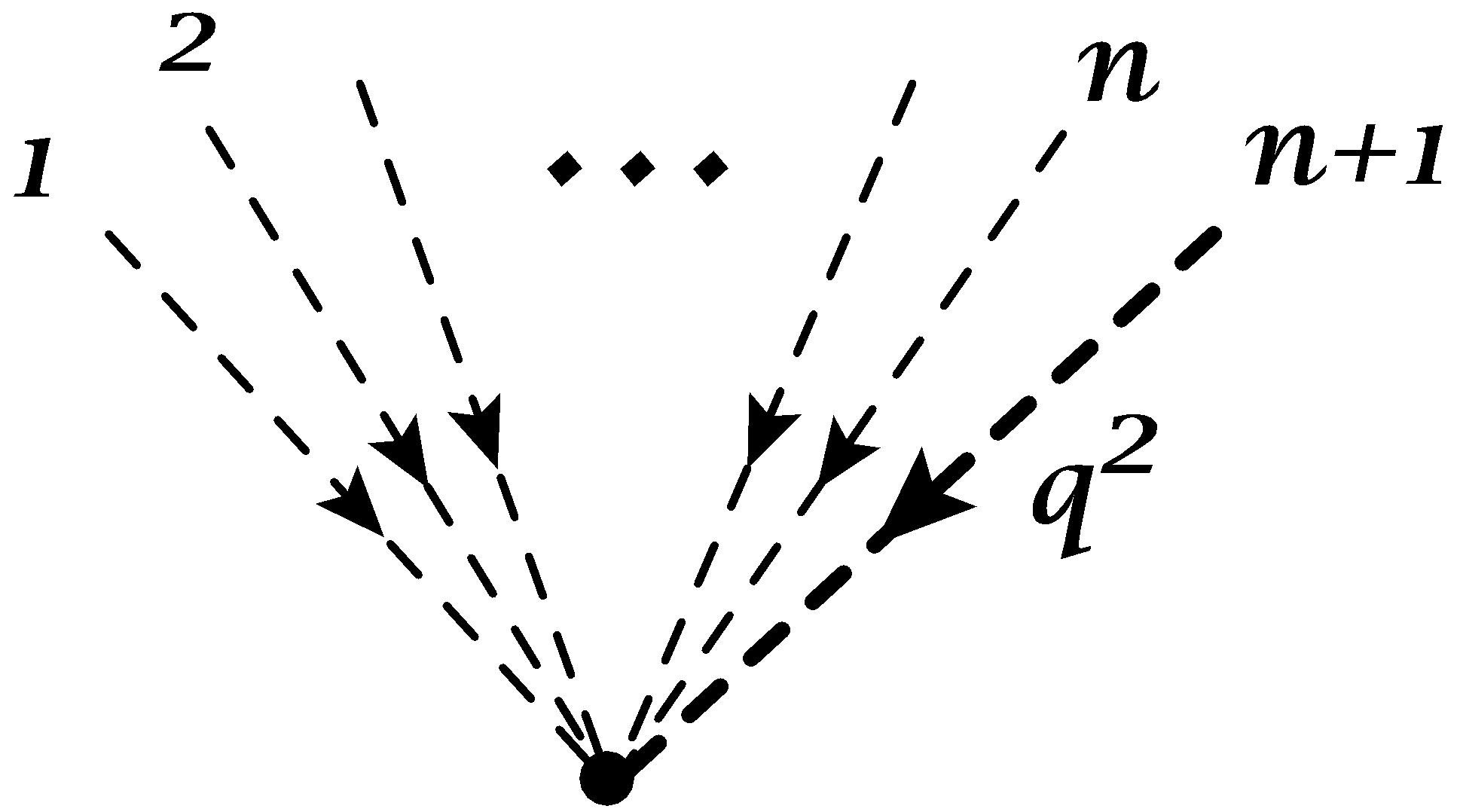}\vspace{1.1cm}
                \caption{$\displaystyle - i c_n g q^2$}
                \label{fig1:frc}
        \end{subfigure}
~\quad           
        \begin{subfigure}[b]{0.2\textwidth}
                \includegraphics[width=\textwidth]{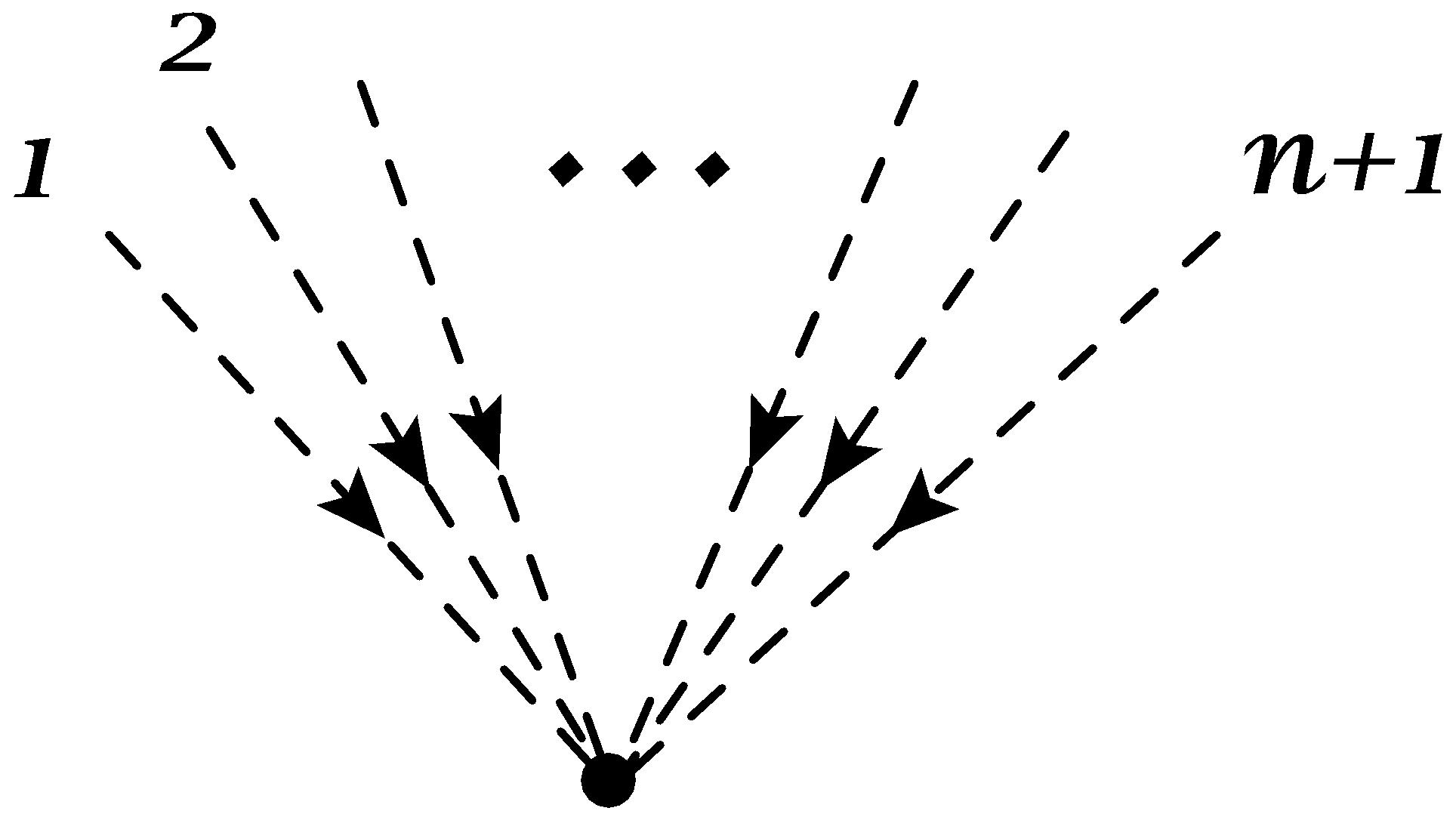}\vspace{1.1cm}
                \caption{$\displaystyle -i m^2 c_n$}
                \label{fig1:frd}
        \end{subfigure}
~\quad     
        \begin{subfigure}[b]{0.2\textwidth}
                \includegraphics[width=\textwidth]{drawings/ScalarRelVert.jpg}\vspace{0.8cm}
                \caption{$\displaystyle -\frac{i}{4}c_k c_{n+1-k}$}
                \label{fig1:fra}
        \end{subfigure}
        \caption{Feynman rules for the relevant interactions. Dashed lines denote bosons, solid lines denote fermions. A thick dashed line represents a boson with a spatial momentum insertion. There is a trace and a factor of $-1$ for a closed fermionic loop. For the bosonic legs there is also a symmetry factor. $n \geq 2$ for vertices (a) through (c) and $n \geq 3$ for vertex (d).}
        \label{fig:FynRelev}
\end{figure}

In this subsection we will study the quantum corrections to the set of interactions described by the Lagrangian \eqref{eq:c2}. The bosonic and fermionic propagators are given by:
\begin{equation}\label{eq:propagators}
\left\langle {\phi \phi } \right\rangle  = \frac{i}{2}\, \frac{1}{{{\omega ^2} - {{\left( {{g k^2} + {m^2}} \right)}^2+i\epsilon}}}
\qquad
\qquad
\left\langle {\psi^\alpha \bar \psi_\beta } \right\rangle  = \frac{1}{2} \, \frac{ \omega\,
(\gamma ^0) {}^\alpha{}_\beta    + i\left( {{g k^2} + {m^2}} \right)\delta^\alpha_\beta} {{{\omega ^2} - {{\left( {{g k^2} + {m^2}} \right)}^2 +i\epsilon}}},
\end{equation}
where the half factor in the fermion propagator is due to the fact that we are working with real fermions.
The relevant Feynman rules for the vertices are summarized in figure \ref{fig:FynRelev}.

\subsubsection{Cancellation of Quadratic Divergences at First Order}\label{subsec:cancellation_1st_order}

\begin{figure}[ht!]
\includegraphics[width=150mm]{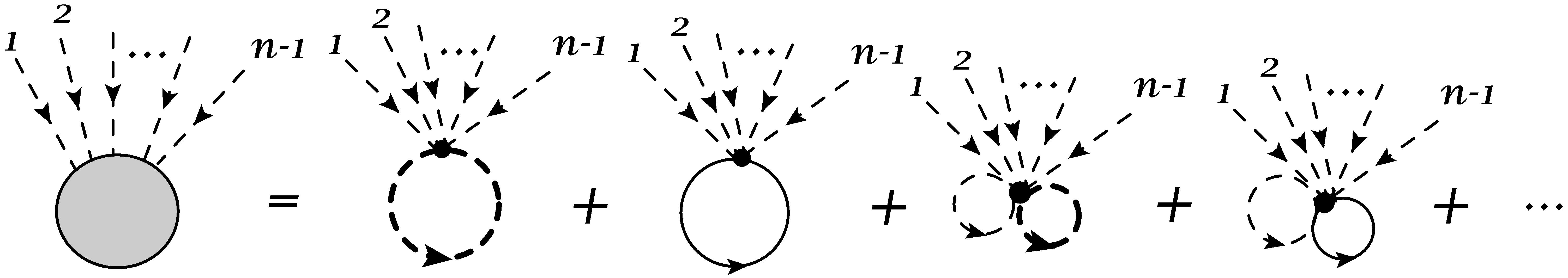}
\caption{First order contributions to a diagram with $n-1$ external bosonic legs and no fermionic legs. Note that the purely bosonic vertices contain momentum insertions (see the vertex in figure \ref{fig1:frc}). \label{fig:cancelation_rel_1}}
\end{figure}

We begin by studying the divergences of the theory at first order in perturbation theory, which is linear in the coupling constants $c_n$ with $n\neq 1$ ($m^2$ is not necessarily small).\footnote{Having more than one coupling constant allows for different perturbative expansions based on different hierarchies of the coupling constants that one can impose when taking their limit to zero. It is an interesting question whether imposing a hierarchy of the coupling constants which makes the infinite series of interactions in equation \eqref{eq:c2} converge for any value of the fields will keep this property at any order in perturbations theory. We can alternatively cut the infinite series of interactions at a finite $n$ in which case higher order interactions will be formed successively at each order in perturbation theory. Formally speaking the limit that we take in this section can be thought of as a strong suppression of coupling constants $c_n$ for all $n>N$ where $N$ is some large integer.}
Any non-vanishing diagram in the theory must have an even number of fermionic external legs. At first order in perturbation theory we have at most 2 external fermionic legs. Such diagrams are at most logarithmically divergent. A diagram with only bosonic external legs can however have quadratically divergent contributions. This is easily seen by dimensional analysis. Here we show that these divergences cancel and we are left with only logarithmically divergent contributions.
All the quadratically divergent contributions to a diagram with $n-1$ external bosonic legs and no external fermionic legs are depicted in figure \ref{fig:cancelation_rel_1}. This includes the corrections to the bosonic propagator for $n=3$.
The quadratic divergences cancel in pairs.

Let us first examine the first pair of diagrams in figure \ref{fig:cancelation_rel_1}. The one with a single bosonic loop has the following quadratically divergent contribution:
\begin{equation}
\label{quant_rel:scal1}
\mathcal{B} =  \frac{c_n  n! }{2}\int \frac{{d{\omega _q}{d^2}q}}{{\left( {2\pi } \right)}^3}\,
\frac{g\, q^2}{{\omega_q}^2 - \left(g\, q^2 + m^2 \right)^2}.
\end{equation}
The frequency integral is convergent and  can be performed explicitly. The natural thing is then to regard the quadratic divergence as a quadratic dependence on the spatial momentum cutoff $\Lambda$.
The quadratically divergent contribution from the fermionic vertex reads:
\begin{equation}
\begin{split}
\label{quant_rel:ferm1}
\mathcal{F}& =
 \frac{i c_n n!}{4} \int \frac{d\omega _q d^2q}{(2\pi)^3} \, \frac{ \tr\left[\gamma ^0 \omega _q + i(g \, q^2 + m^2)\right]} {{\omega _q}^2 - (g\, q^2 + m^2)^2}
 \\&
 =
 -\frac{c_n n!}{2}
 \int \frac{d\omega _q d^2 q}{(2\pi)^3}
 \frac{g\, q^2}{{\omega_q}^2 - (g\, q^2 + m^2 )^2} + \dots,
 \end{split}
\end{equation}
where in the last step we have neglected logarithmically divergent contributions and used $\tr \gamma^0=0$ and $\tr \delta^\alpha_\beta = 2$.
When summing these two diagrams one finds that the quadratic divergence cancels. Hence $\mathcal B+\mathcal F $ is at most logarithmically divergent. 

The following pairs of diagrams in figure \ref{fig:cancelation_rel_1} would in general contain in addition to the structure described above $k$ more closed bosonic loops without momentum insertions. Let us denote the standard (logarithmically divergent) contribution of a bosonic loop with no momentum insertions by:
\begin{equation}\label{QuantRel_BL}
\mathcal{BL} \equiv \int \frac{ d\omega_q d^2q}{(2\pi)^3}
\, \frac{i}{2} \, \frac{1}{\omega^2-(g \, q^2+m^2)^2} = \frac{1}{16\pi g }\log\left(\frac{g \Lambda^2}{m^2}\right)+\text{finite}.
\end{equation}
Then the $c_n n!$ factor in \eqref{quant_rel:scal1}-\eqref{quant_rel:ferm1} would be replaced by:
\begin{equation}
\frac{(n+2k)!c_{n+2k}}{2^k k!} (\mathcal{BL})^k ,
\end{equation}
for each of these pairs of diagrams.
This factor is common to the fermionic and bosonic diagrams for each pair and hence does not change the conclusion about the cancellation of quadratic divergences in each pair of diagrams.

\subsubsection{Renormalized First Order Perturbation Theory}
\label{subsec:RenFirsPertTh}
\begin{figure}
   \centering
		\subcaptionbox{$\displaystyle  \frac{i}{2}\, \frac{1 \phantom{{\gamma^0}^\alpha{}_\beta}}{{{\omega ^2} - {{\left( g \, k^2 + m^2 \right)}^2+i\epsilon}}}$                \label{fig:frr_scalprop}}
		{\includegraphics[width=0.36\textwidth]{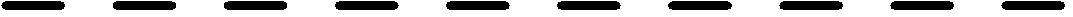}}
	\qquad \qquad
		\subcaptionbox{$\displaystyle \frac{1}{2} \, \frac{ \omega\,
(\gamma ^0) {}^\alpha{}_\beta    + i\left( g\,k^2 + {m^2} \right)\delta^\alpha_\beta} {{{\omega ^2} - {{\left( g \, k^2 + {m^2} \right)}^2 +i\epsilon}}}$ \label{fig:frr_fermprop}}
		{\includegraphics[width=0.38\textwidth]{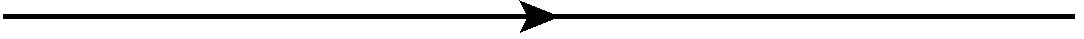}}
\\
\vspace{1cm}
   \centering
		\subcaptionbox{$\displaystyle i\delta_{z_\phi} (\omega^2-g^2 p^4+m^4)
		\\ ~~~~-2i(\delta_m+p^2\delta_g) (m^2+g p^2) $                \label{fig:frr_scalprop2}}
		{\includegraphics[width=0.38\textwidth]{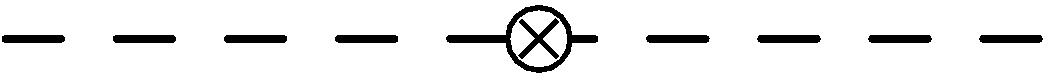}}
	\qquad \qquad
		\subcaptionbox{$\displaystyle \delta_{z_\psi}
		(\omega\gamma^0 - i(g p^2+m^2))
		\\
		~~~ ~~~ +i m^2\delta_{z_\phi} -i(\delta_m + p^2 \delta_g) $ \label{fig:frr_fermprop2}}
		{\includegraphics[width=0.39\textwidth]{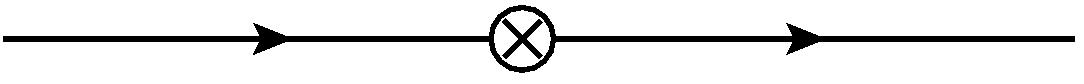}}
\\
\vspace{0.5cm}              
        \begin{subfigure}[b]{0.25\textwidth}
                \includegraphics[width=\textwidth]{drawings/ScalarRelVert.jpg}\\ \\
                \caption{$\displaystyle -i m^2 c_n$}
                \label{fig:frra}
        \end{subfigure}
~\quad        
        \begin{subfigure}[b]{0.25\textwidth}
                \includegraphics[width=\textwidth]{drawings/FermionRelVert.jpg}
                \caption{$\displaystyle - \frac{i}{2} n c_n $}
                \label{fig:frrb}
        \end{subfigure}
~\quad
        \begin{subfigure}[b]{0.25\textwidth}
                \includegraphics[width=\textwidth]{drawings/ScalarRelVertMom.jpg}\\ \\
                \caption{$\displaystyle - i c_n\, g\, q^2$}
                \label{fig:frrc}
        \end{subfigure}%
\\
\vspace{1cm}              
        \begin{subfigure}[b]{0.25\textwidth}
                \includegraphics[width=\textwidth]{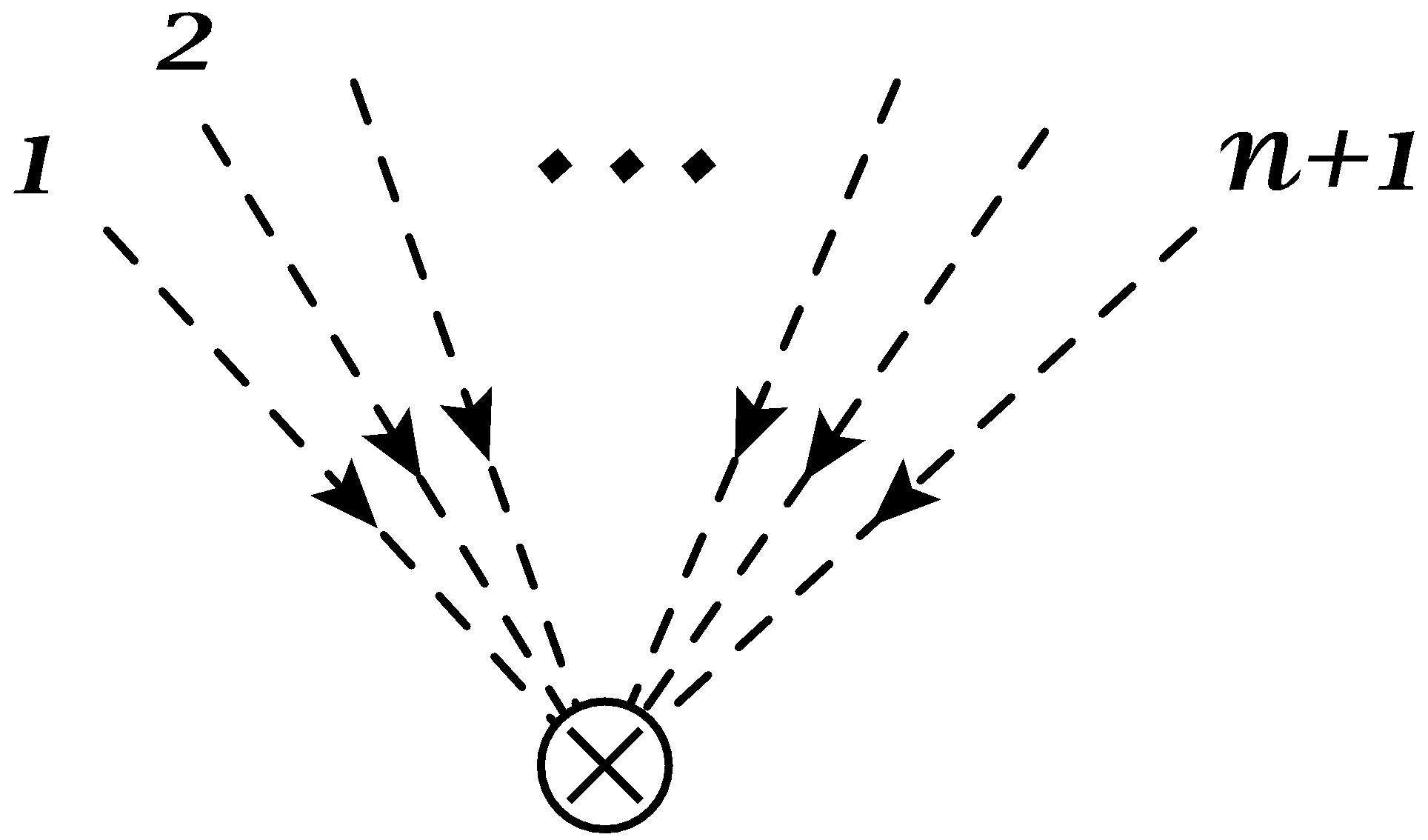}\\ \\
                \caption{$\displaystyle -i m^2 \delta_{c_n}$}
                \label{fig:frrda}
        \end{subfigure}
~\quad        
        \begin{subfigure}[b]{0.25\textwidth}
                \includegraphics[width=\textwidth]{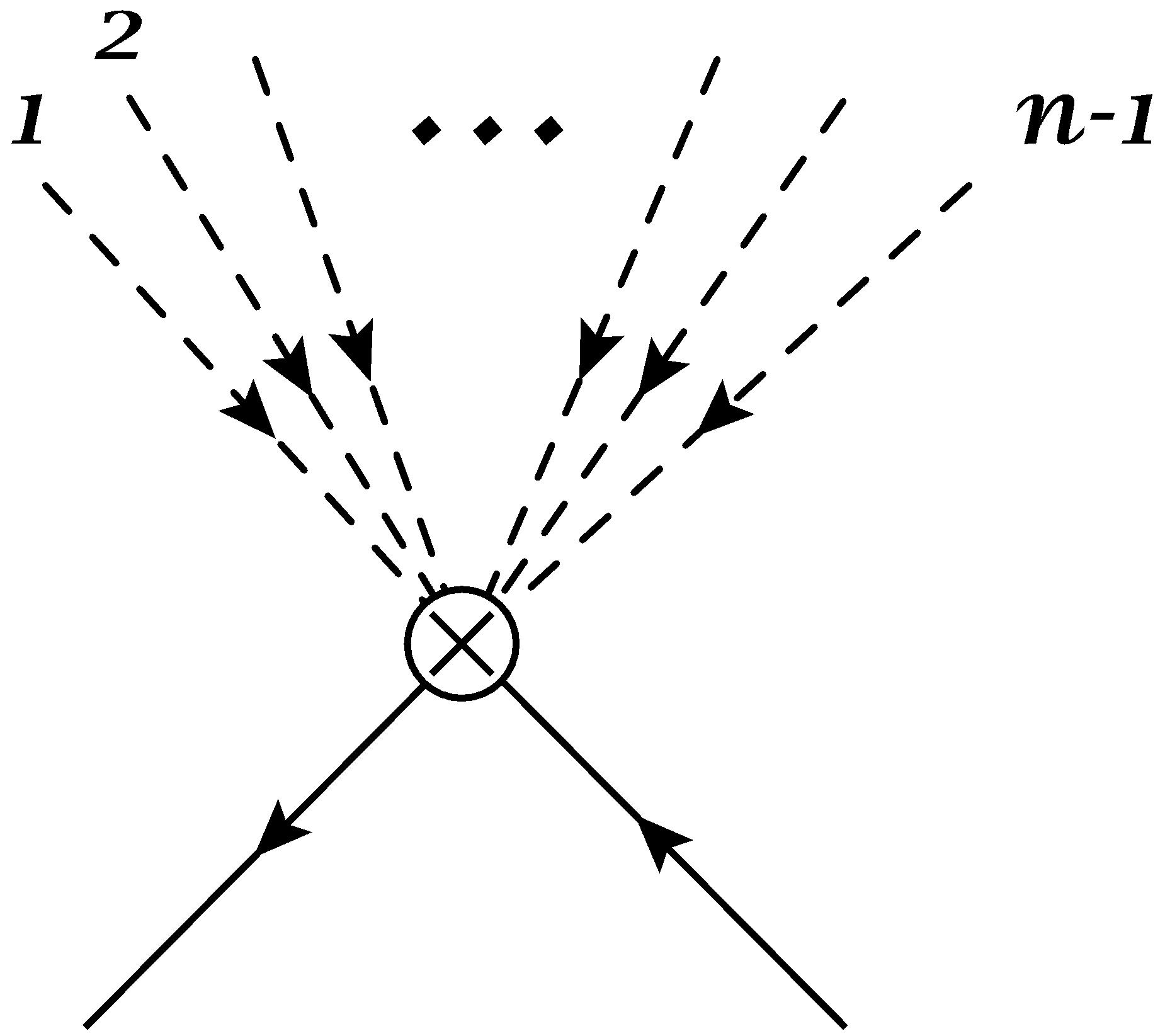}
                \caption{$\displaystyle - \frac{i}{2} n \delta_{c_n} $}
                \label{fig:frrdb}
        \end{subfigure}
~\quad
        \begin{subfigure}[b]{0.25\textwidth}
                \includegraphics[width=\textwidth]{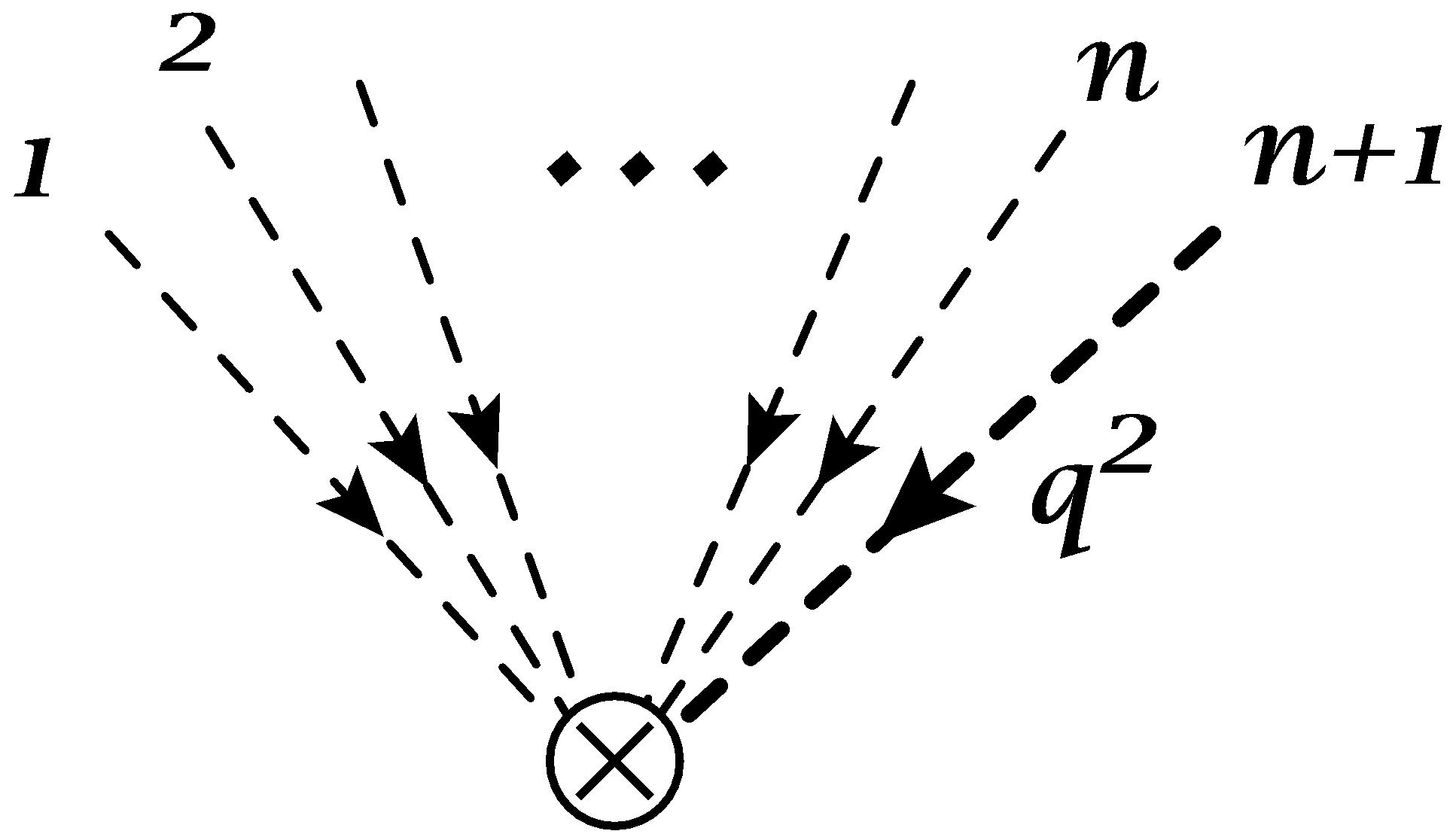}\\ \\
                \caption{$\displaystyle - i \delta_{c_n} \, g\, q^2$}                \label{fig:frrdc}
          \end{subfigure}              
        \caption{Feynman rules for the relevant interactions in renormalized perturbation theory. We kept only those contributions which are at most linear in the coupling constants.}
        \label{fig:FynRelev2}
\end{figure}

We need to absorb the non-physical logarithmic divergences  that remain in the theory at first order in the perturbative expansion into the unobservable bare parameters of the theory defined by:
\begin{equation}\label{eq:relevant_interactions_revisited2}
\begin{split}
\mathcal L^c = &
 \left( \partial _t \phi  \right)^2
  -  g_0^2 \left( \nabla ^2 \phi  \right)^2 
   + 2m_0^2 \, g_0 \, \phi \nabla^2\phi
   - m_0^4 \phi^2 
   +  \bar \psi \gamma ^0  \partial _t \psi
   + g_0\, \bar \psi \nabla ^2\psi
   - m_0^2 \bar \psi \psi 
   \\&
       -   \sum\limits_{n,m = 2}^\infty  \frac{c^0_n c^0_m}{4} \phi ^{n+m}  
   - m_0^2  \sum\limits_{n = 2}^\infty  c^0_n \phi ^{n+1} 
  -  \sum\limits_{n = 2}^\infty  \frac{c^0_n n}{2} \phi ^{n - 1} \bar \psi \psi 
 + \sum\limits_{n = 2}^\infty g_0\, c^0_n  \phi^n \nabla^2 \phi,
 \end{split}
\end{equation}
where we rewrote the mass, relative normalization of the kinetic terms and coupling constants as $m_0, g_0, c^0_n$ to emphasize that these are the bare parameters of the theory.
Redefining:
\begin{equation}\label{eq:RedefRelParams}
\begin{split}
&\phi = Z_\phi^{1/2} \phi_r, \qquad 
\psi = Z_\psi^{1/2} \psi_r,\\
& \delta_{z_\phi} =  Z_\phi-1, \qquad
\delta_{z_\psi} = Z_\psi-1,
\\&
\delta_{m} = m_0^2 Z_\phi - m^2, \qquad
\delta_g = g_0 - g,
\\&
\delta_{c_n} = c^0_n Z_\phi^{(n-1)/2} - c_n,
\end{split}
\end{equation}
where now $m, c_n$ are the physically measured mass and coupling constants, we can recast the Lagrangian in the form:
\begin{equation}\label{QC:RelL1Ord}
\begin{split}
\mathcal L^c = & \,
     (1+\delta_{z_\phi}) \left( \partial _t \phi_r  \right)^2
  -  g (g(1+\delta_{z_\phi})+2\delta_g)  \left( \nabla ^2 \phi_r  \right)^2    
\\&   
    + 2 (g(m^2 + \delta_{m})+m^2\delta_g)  \phi_r \nabla^2\phi_r - (m^4(1-\delta_{z_\phi}) + 2m^2\delta_{m})  \phi_r^2 
   \\&
   + (1+\delta_{z_\psi}) \bar \psi_r \gamma ^0  \partial _t \psi_r
   + (g(1+\delta_{z_\psi})+\delta_g) \bar \psi_r \nabla ^2\psi_r
   \\&
   - (m^2  (1+\delta_{z_\psi}-\delta_{z_\phi})  + \delta_{m}) \bar \psi_r \psi_r 
   \\&
   - (c_n+ \delta_{c_n}) \sum\limits_{n = 2}^\infty  
   \phi_r ^{n - 1} \left[ m^2\phi_r ^2    
  -g   \phi_r \nabla^2 \phi_r
    +  \frac{ n}{2}  \bar \psi_r \psi_r \right] + O(c_{n>1}^2,\delta^2),
 \end{split}
\end{equation}
which corresponds to the Feynman rules in figure \ref{fig:FynRelev2}. We will only study the first non-trivial corrections to the field strengths, relative normalization of the kinetic terms, mass and coupling constants. Because vertices can produce closed loops that emerge from them the first non-trivial corrections to the coupling constants are of the same order as the coupling constants themselves.

In addition to the redefinitions \eqref{eq:RedefRelParams} we need to specify the precise definitions of the physical mass and coupling constants. The mass and field normalization are defined by requiring that the poles in the propagators and their residues remain the same as in the non-corrected propagator \eqref{eq:propagators}. Since the diagrams contributing to the corrections to the bosonic and fermionic propagator at first order in the coupling constants do not in general depend on the external frequency this forces $\delta_{z_\phi} = \delta_{z_\psi} = 0$. 

\begin{figure}[ht!]
\centering
\includegraphics[width=\textwidth]{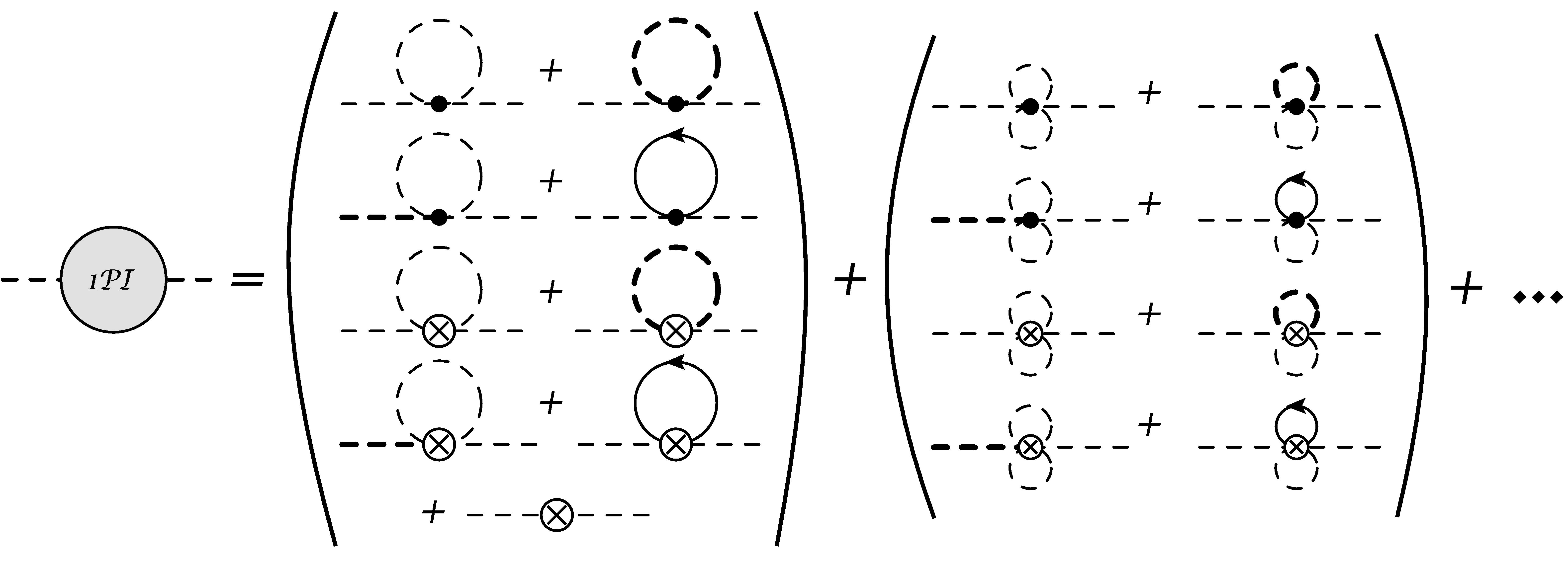}
\caption{Corrections to the bosonic propagator. \label{fig:ScalarPropCorrectionsRel}}
\end{figure}

\begin{figure}[ht!]
\centering
\includegraphics[width=\textwidth]{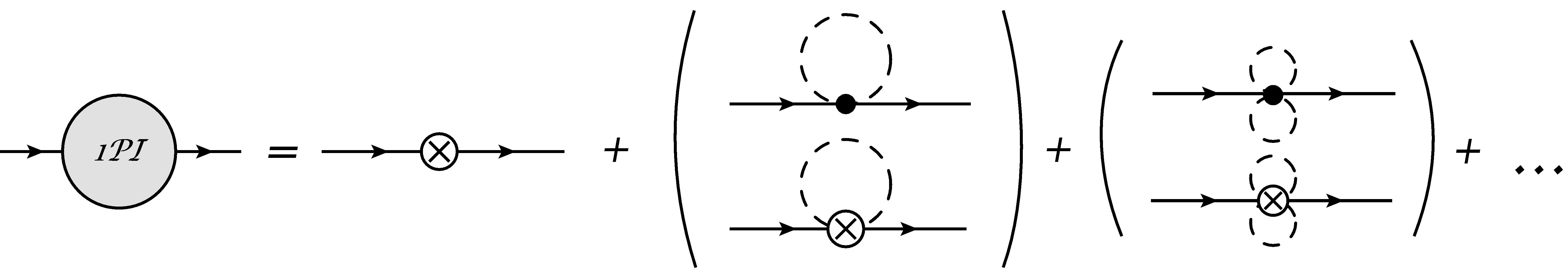}
\caption{Corrections to the fermionic propagator. \label{fig:FermionPropCorrectionsRel}}
\end{figure}

The first order perturbative corrections to the bosonic and fermionic propagators are depicted in figures \ref{fig:ScalarPropCorrectionsRel} and \ref{fig:FermionPropCorrectionsRel}.
If we again denote:
\begin{equation}\label{QuantRel_BL_full}
\mathcal{BL} \equiv \int \frac{ d\omega_q d^2q}{(2\pi)^3}
\, \frac{i}{2} \, \frac{1}{{\omega_q}^2-(g \, q^2+m^2)^2} =  \frac{1}{16\pi g }\log\left(\frac{g \Lambda^2+m^2}{m^2}\right),
\end{equation}
we can write the result of figure \ref{fig:ScalarPropCorrectionsRel} as follows:
\begin{equation}\label{D_phi}
\mathcal{D}_\phi = -i ( g p^2+m^2) \left[4 (\delta_m +p^2\delta_g) + \sum\limits_{k=1}^\infty  \frac{(2k+1)!}{2^{k-1} k!} \left(c_{2k+1}+\delta_{c_{2k+1}}\right) \cdot  (\mathcal{BL})^k \right],
\end{equation}
where $p$ is the spatial external momentum, and that of figure \ref{fig:FermionPropCorrectionsRel} as:
\begin{equation}\label{D_psi}
\mathcal{D}_\psi = -i \left[\delta_m +p^2 \delta_g + \sum\limits_{k=1}^\infty \frac{(2k+1)!}{2^{k+1} k!} \left(c_{2k+1}+\delta_{c_{2k+1}}\right) \cdot  (\mathcal{BL})^k \right].
\end{equation}
Setting $\delta_g=0$ and:
\begin{equation}
\delta_m = - \sum\limits_{k=1}^\infty  \frac{(2k+1)!}{2^{k+1} k!} \left(c_{2k+1}+\delta_{c_{2k+1}}\right) \cdot  (\mathcal{BL})^k
\end{equation}
yields $\mathcal{D}_\phi=\mathcal{D}_\psi=0$ for all $p^2$. This
is consistent with the conditions on the propagators' poles and residues. The physical definition of $\delta_{c_n}$ will be detailed below.
Note that the fact that the corrections to the three independent terms $\phi\nabla^2\phi$, $\phi^2$ and $\bar\psi \psi$ can all be reabsorbed in the same $\delta_m$ is highly non-trivial and indicates the quantum conservation of supersymmetry at first order in perturbation theory.
We will prove a similar conclusion for $\delta_{c_n}$ in what follows. 

\begin{figure}[ht!]
\centering
\includegraphics[width=\textwidth]{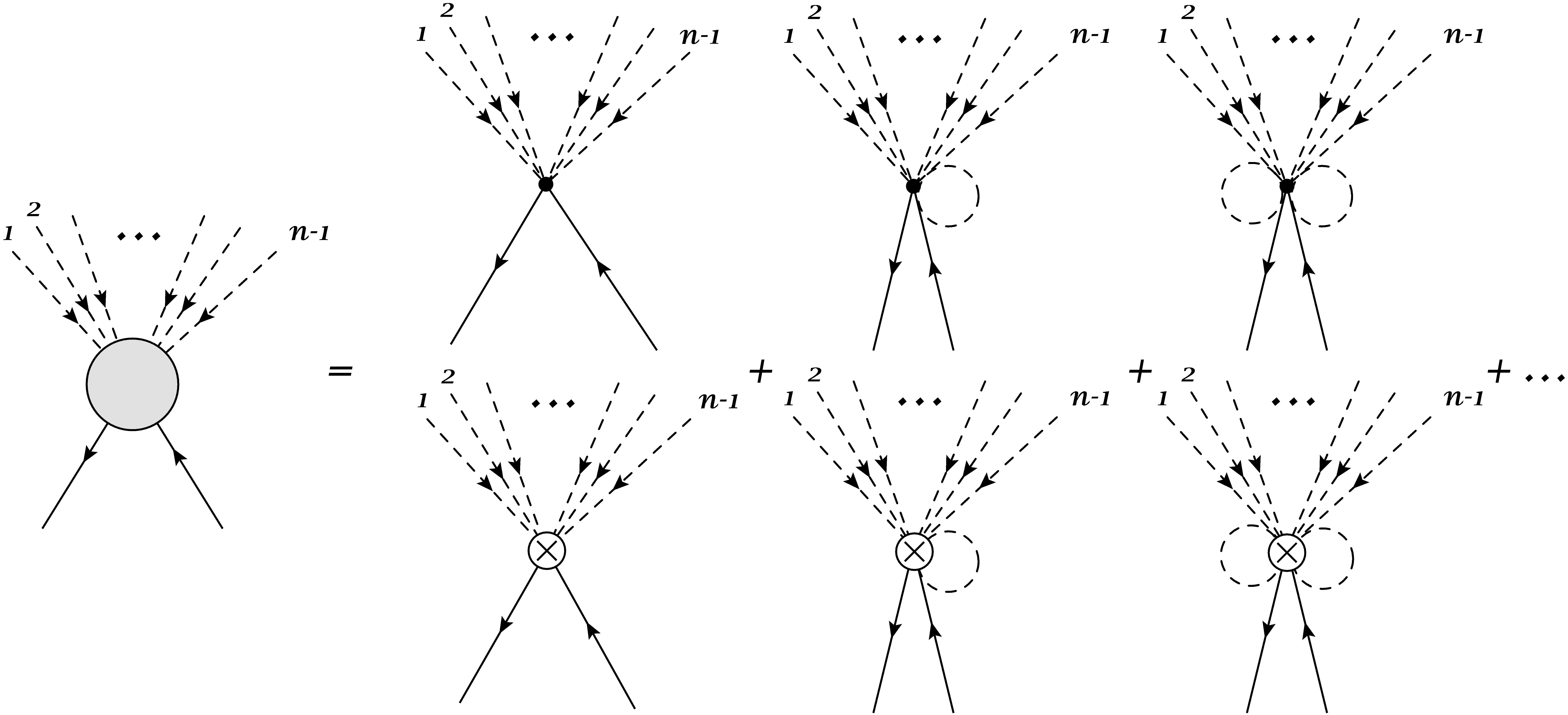}
\caption{Corrections to a vertex with 2 fermionic and $n-1$ bosonic legs. \label{fig:FermionRelVertCorrections}}
\end{figure}

\begin{figure}[ht!]
\centering
\includegraphics[width=\textwidth]{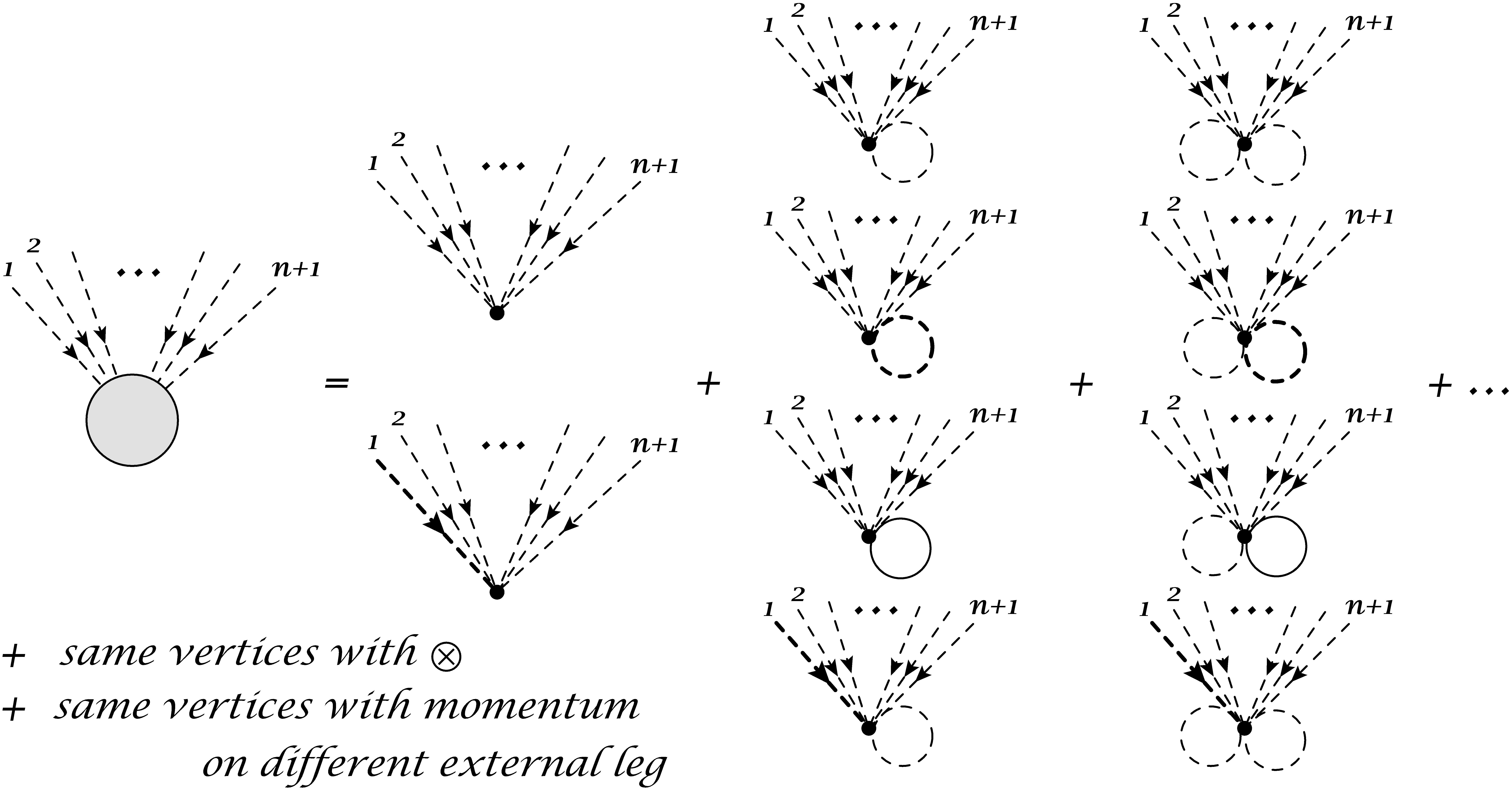}
\caption{Corrections to a vertex with $n+1$ bosonic legs. \label{fig:ScalarRelVertCorrections}}
\end{figure}

We now proceed to the corrections to the vertices. At first order in perturbation theory, a non-vanishing diagram has at most 2 external fermionic legs. The contributions to a scattering amplitude with 2 external fermionic legs and $n-1$ external bosonic legs ($n\geq 2$) are depicted in figure \ref{fig:FermionRelVertCorrections}. The value of the amplitude is given by:
\begin{equation}\label{eq:fermion_blob}
\mathcal{M}_\psi^n = -\frac{i}{2} \left[n!\, \widetilde c_n  + \sum \limits_{k=1}^{\infty} \, \widetilde c_{n+2k} \frac{(n+2k)!}{2^k k!} (\mathcal{BL})^k \right].
\end{equation}
The scattering amplitude with $n+1$ bosonic legs is depicted in figure \ref{fig:ScalarRelVertCorrections} and equals:
\begin{equation}\label{eq:scalar_blob}
\mathcal{M}_\phi^n = - i \left(m^2 (n+1)+ \sum\limits_{ i \in bosonic\atop external\ legs} g p_{\phi,i}^2 \right) \left[ n! \, \widetilde c_n  + \sum \limits_{k=1}^{\infty}\, \widetilde c_{n+2k} \frac{(n+2k)!}{2^k k!} (\mathcal{BL})^k \right],
\end{equation}
where we have defined $\widetilde c_n = c_n+\delta_{c_n}$.

To make an appropriate choice of the corrections one needs to properly define the coupling constants. A frequent requirement is that the appropriate amplitude equals the tree level value (no-loops) of the relevant coupling constant for certain values of the external momenta. This is obtained by for example choosing:
\begin{equation}\label{c_n_corr}
\delta_{c_n}  = - \frac{1}{n!}\sum \limits_{k=1}^{\infty}\, \widetilde c_{n+2k} \frac{(n+2k)!}{2^k k!} (\mathcal{BL})^k,
\end{equation}
which will keep the structure of the original vertices for any value of the external momentum. This is due to the fact that the loop integrals do not depend on the external momentum at this order.	
Note also that the corrections for a certain $c_n$ depend on all higher $c_n$ couplings. As mentioned in a previous footnote, we regard the infinite series of interactions as effectively cutoff at some large $N$ so that the infinite sum of corrections is in fact finite and convergent. Other choices in which a precise hierarchy of coupling constants is enforced are also possible (for example a Sine-Gordon like model).

Note again that the fact that the independent corrections to the three terms $m  c_n \phi ^{n+1}$, $c_n  \phi^n \nabla^2 \phi$ and $c_n \phi ^{n - 1} \bar \psi \psi$ can all be reabsorbed in the same set of corrections $\delta_{c_n}$ is  highly non-trivial, and indicates that supersymmetry is preserved at  first order in perturbation theory.

\subsubsection{Cancellation of Quadratic Divergences at Second Order}\label{subsec:Cancellation_2nd_order}
The absence of quadratic divergences persists at second order in perturbation theory. 
Here we should consider diagrams quadratic in the coupling constants $c_n$. Diagrams consisting of one vertex of the type in figure \ref{fig1:fra} with both $c_n \neq c_1$ are at most logarithmically divergent as they can only contain divergent integrals of the form $\mathcal{BL}^m$, where $m$ is the number of bosonic loops closing on themselves. We therefore  consider diagrams with two vertices of the types in figures \ref{fig:frra}--\ref{fig:frrc} and their first order quantum corrections \ref{fig:frrda}--\ref{fig:frrdc}.
In fact, we can replace each of the two vertices plus the contributions due to loops originating from them and closing on themselves and the counter terms that we added at first order by the effective ``blobs'' of figures \ref{fig:FermionRelVertCorrections} and \ref{fig:ScalarRelVertCorrections} (see also equations \eqref{eq:fermion_blob} and \eqref{eq:scalar_blob}). Each of these ``blobs'' (or dressed vertices) is finite and equals the tree level contribution of the appropriate vertex.
We are then left to consider any number of loops closing between the two ``blobs'' (or dressed vertices).

A point that we have ignored so far is that the counting of the degree of divergence is in general non-trivial and depends on the regularization scheme. The naive degree of divergence is obtained by dimensional analysis and is more precisely defined by imposing a correlated cutoff in frequency and momentum after wick rotating the frequency variables $\omega\rightarrow i \omega$. In the case of Lifshitz with $z=2$ it is natural to cut the momentum integrals at a scale of order $\Lambda$ and the frequency integral at $\Lambda^2$, where $\Lambda$ is a spatial momentum cutoff. One can then study the contributions to the integral from very large frequencies and momenta assuming that the most divergent contributions come from this range. The naive degree of divergence is therefore given as:
\begin{equation}
D = -2P_\psi-4P_\phi+2V_{\nabla^2}+4L,
\end{equation}
where $P_\psi$, $P_\phi$ are the number of fermionic and bosonic propagators, $V_{\nabla^2}$ is the number of bosonic vertices with spatial momentum insertions and $L$ is the number of loops.
An alternative approach is to perform explicitly the frequency integrals when those converge and then impose a cutoff on the resultant momentum integral.
Since Lifshitz theories are not boost invariant there is a priori no reason to assume a correlated regularization scheme on frequency and momentum, and one can impose different unrelated cutoffs in the two.

\begin{figure}
   \centering
\subcaptionbox{\label{fig:Rel2OrdCancel-a}}
		{\includegraphics[width=0.25\textwidth]{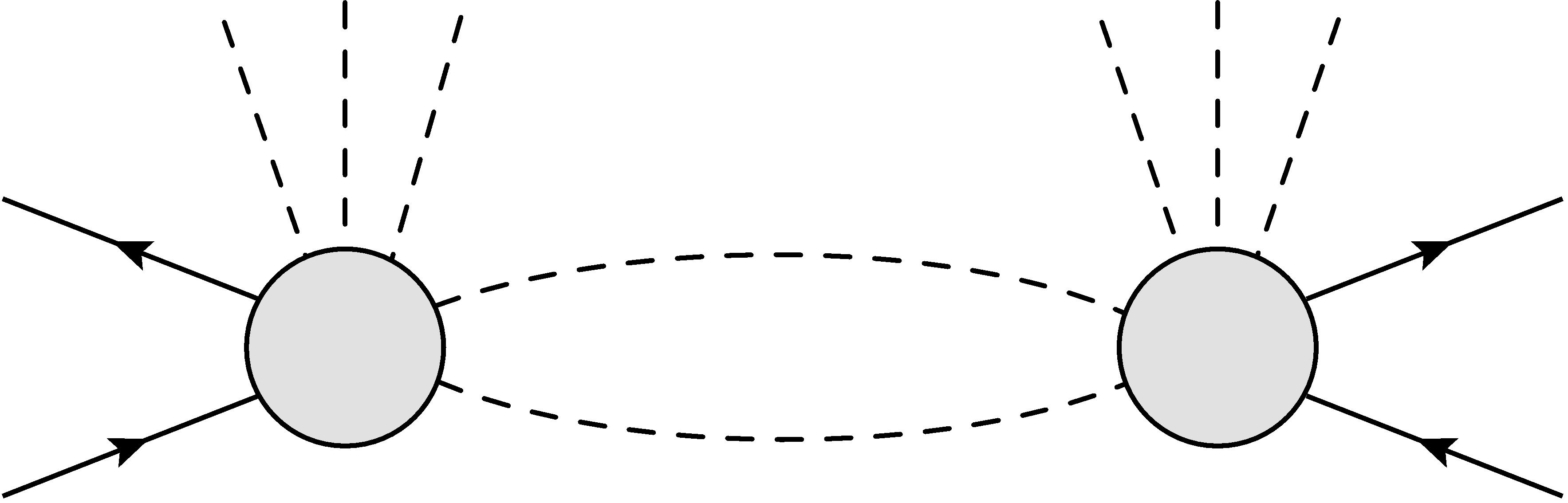}}
	\qquad \qquad
\subcaptionbox{\label{fig:Rel2OrdCancel-b}}
		{\includegraphics[width=0.25\textwidth]{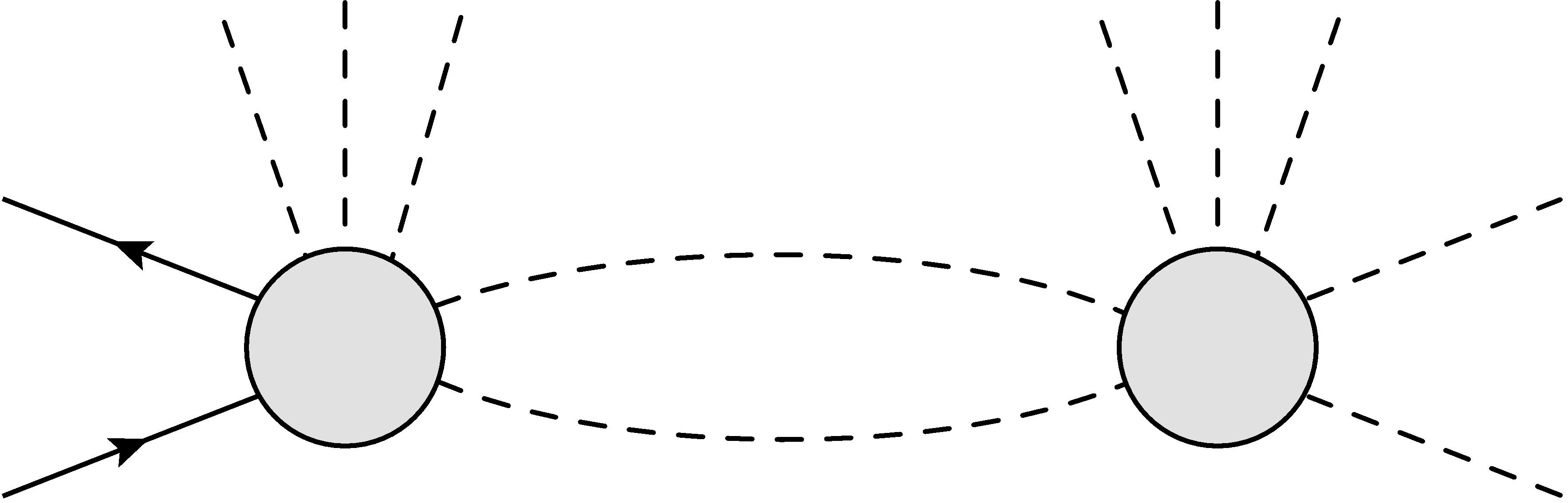}}
			\qquad \qquad
\subcaptionbox{\label{fig:Rel2OrdCancel-c}}
		{\includegraphics[width=0.25\textwidth]{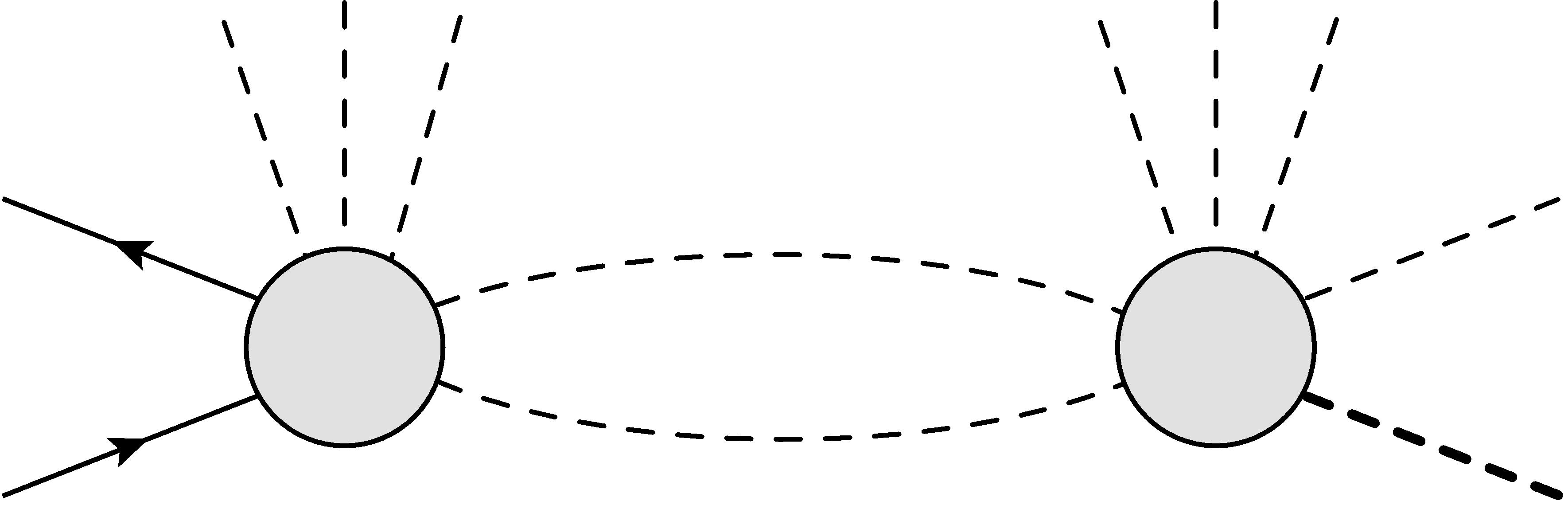}}
\\
\vspace{0.3cm}
   \centering
	\subcaptionbox{\label{fig:Rel2OrdCancel-d}}
		{\includegraphics[width=0.25\textwidth]{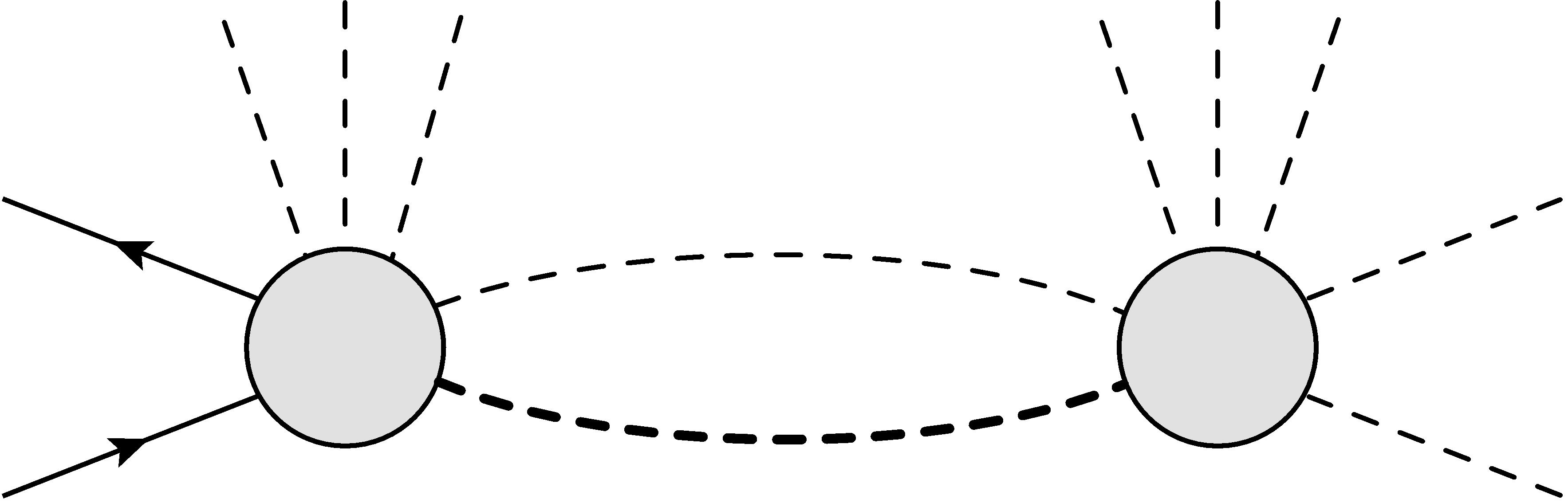}}
\qquad \qquad
		\subcaptionbox{\label{fig:Rel2OrdCancel-e}}
		{\includegraphics[width=0.25\textwidth]{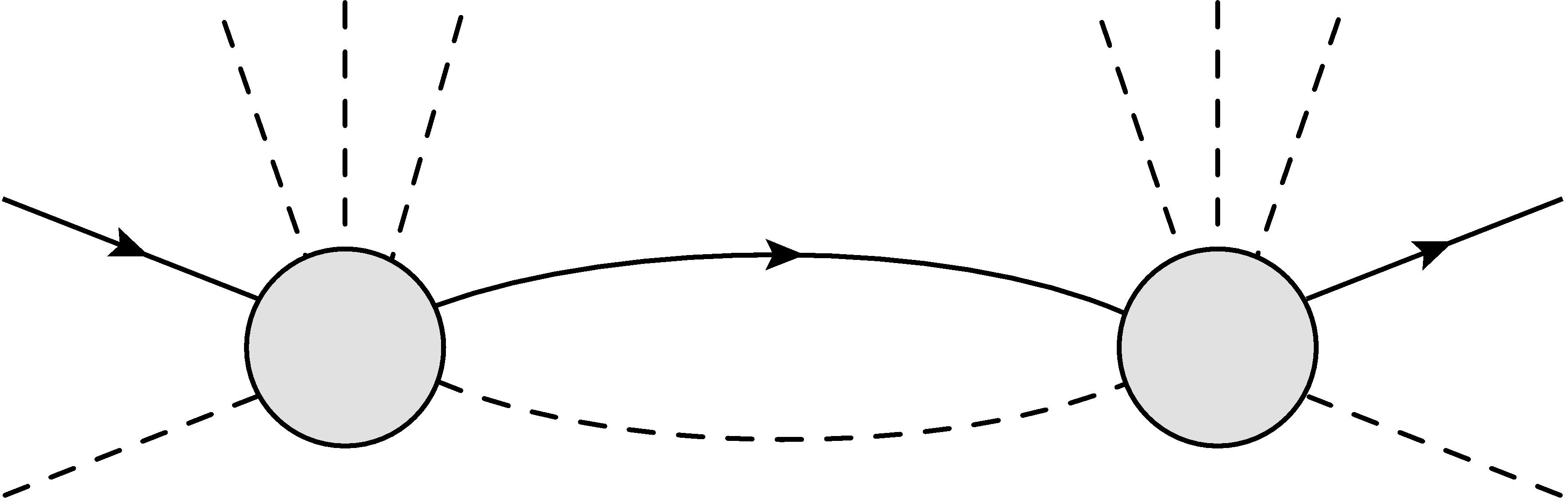}}
\qquad \qquad
	\subcaptionbox{\label{fig:Rel2OrdCancel-f}}
		{\includegraphics[width=0.25\textwidth]{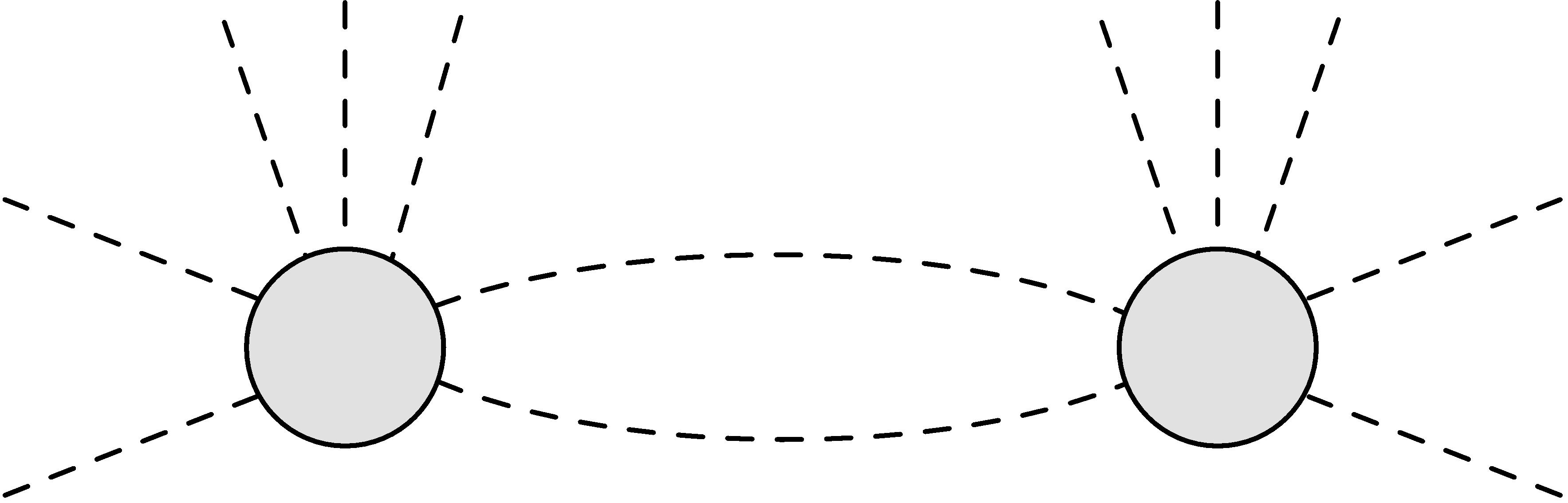}}
\\
\vspace{0.3cm}
   \centering
		\subcaptionbox{\label{fig:Rel2OrdCancel-g}}
		{\includegraphics[width=0.25\textwidth]{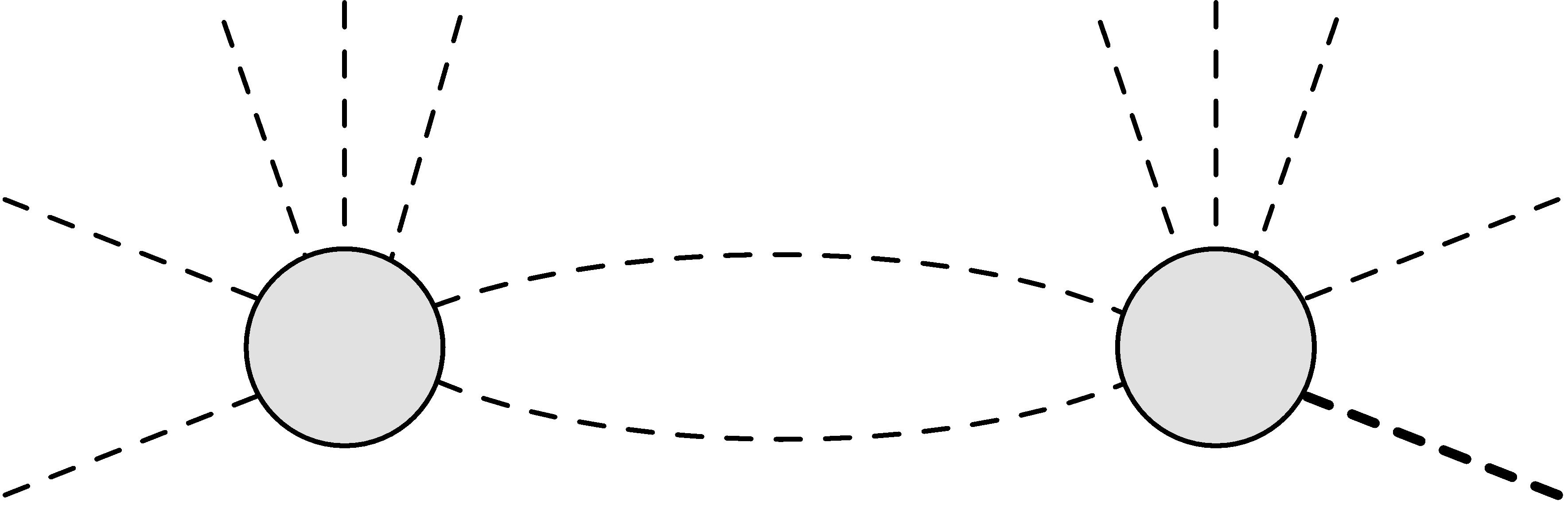}}
\qquad \qquad
	\subcaptionbox{\label{fig:Rel2OrdCancel-h}}
		{\includegraphics[width=0.25\textwidth]{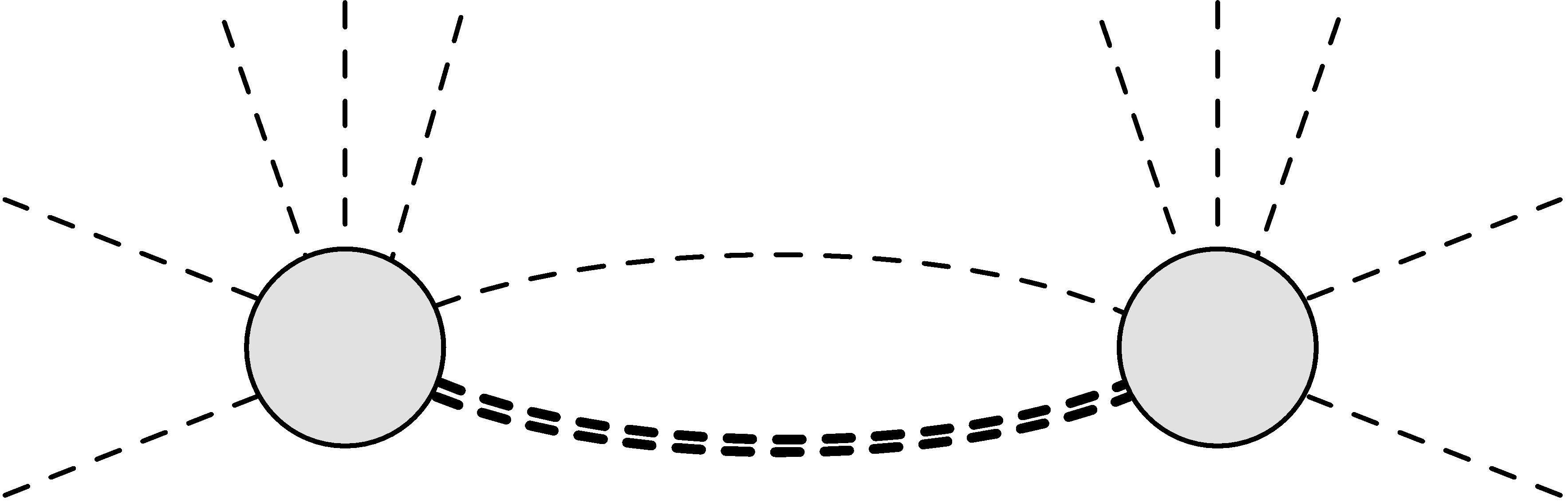}}
\qquad \qquad
		\subcaptionbox{\label{fig:Rel2OrdCancel-i}}
		{\includegraphics[width=0.25\textwidth]{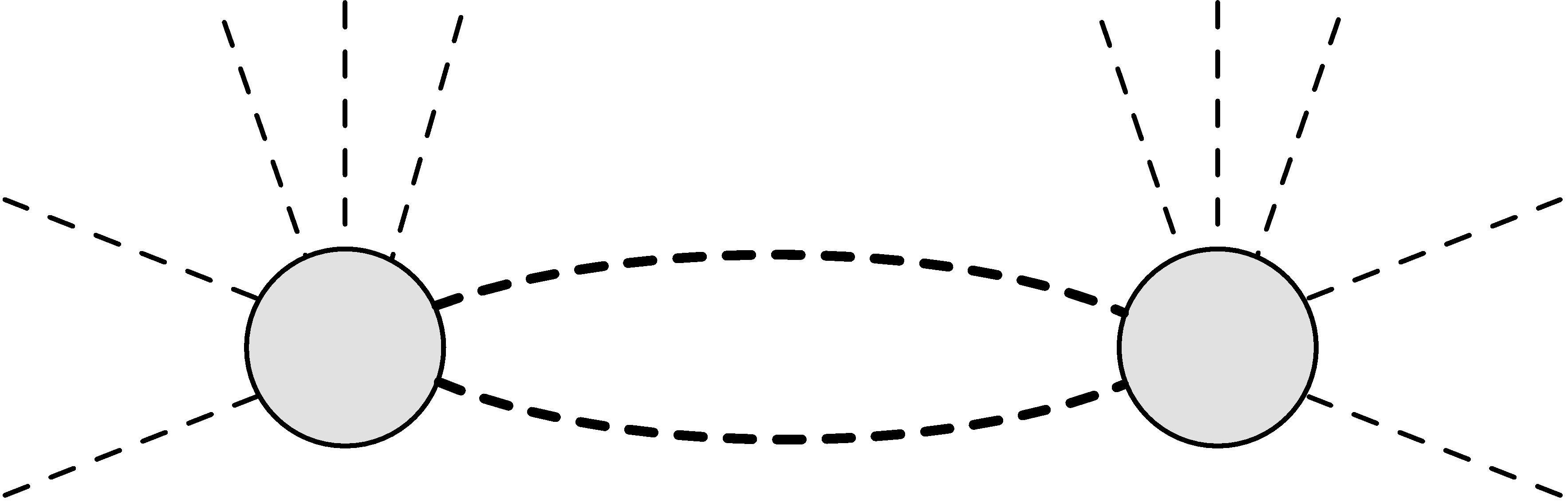}}
\\
\vspace{0.3cm}
   \centering
	\subcaptionbox{\label{fig:Rel2OrdCancel-j}}
		{\includegraphics[width=0.22\textwidth]{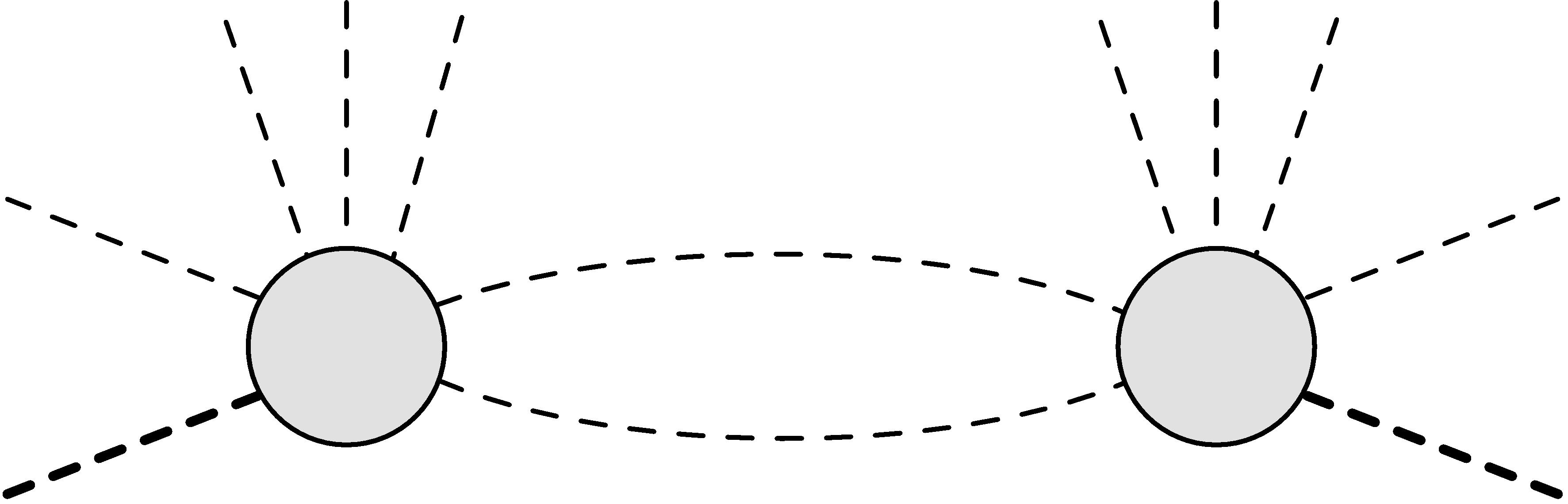}}
\quad
\subcaptionbox{\label{fig:Rel2OrdCancel-k}}
		{\includegraphics[width=0.22\textwidth]{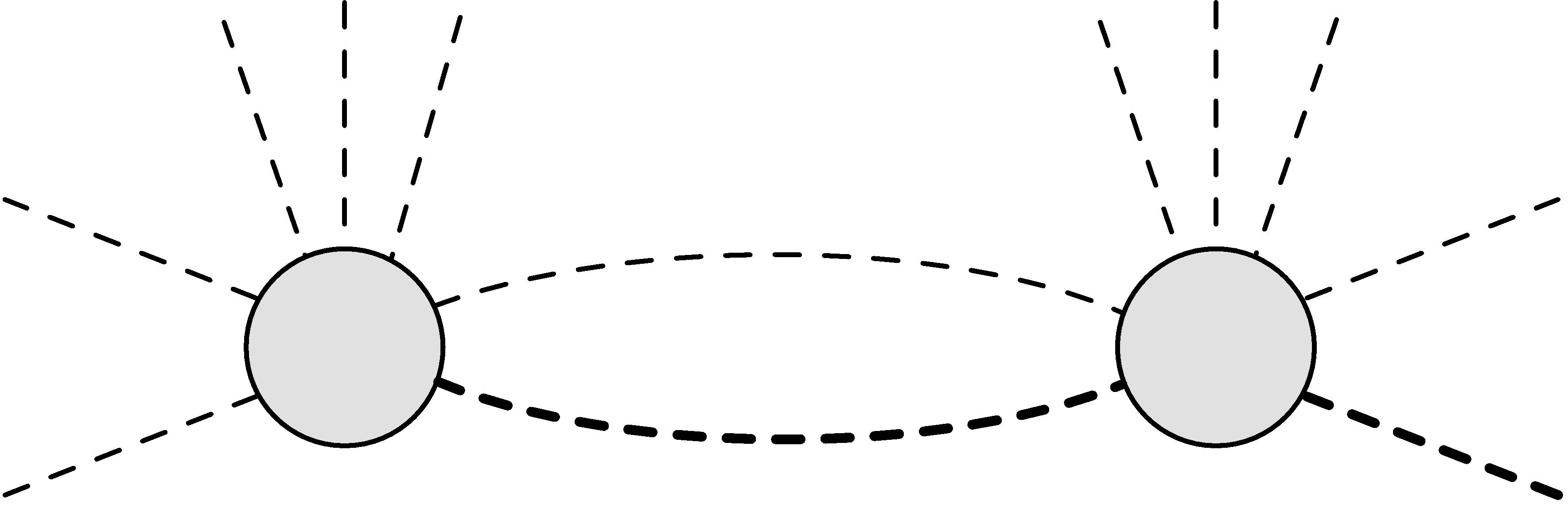}}
\quad
	\subcaptionbox{\label{fig:Rel2OrdCancel-l}}
		{\includegraphics[width=0.22\textwidth]{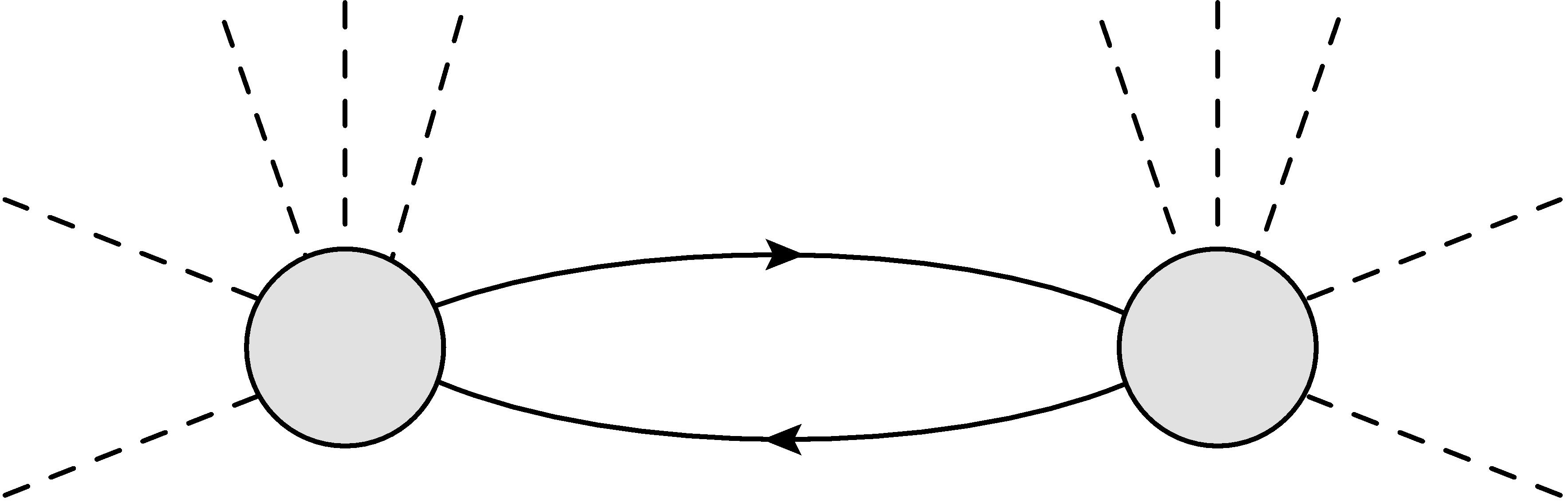}}
\quad
	\subcaptionbox{\label{fig:Rel2OrdCancel-m}}
		{\includegraphics[width=0.22\textwidth]{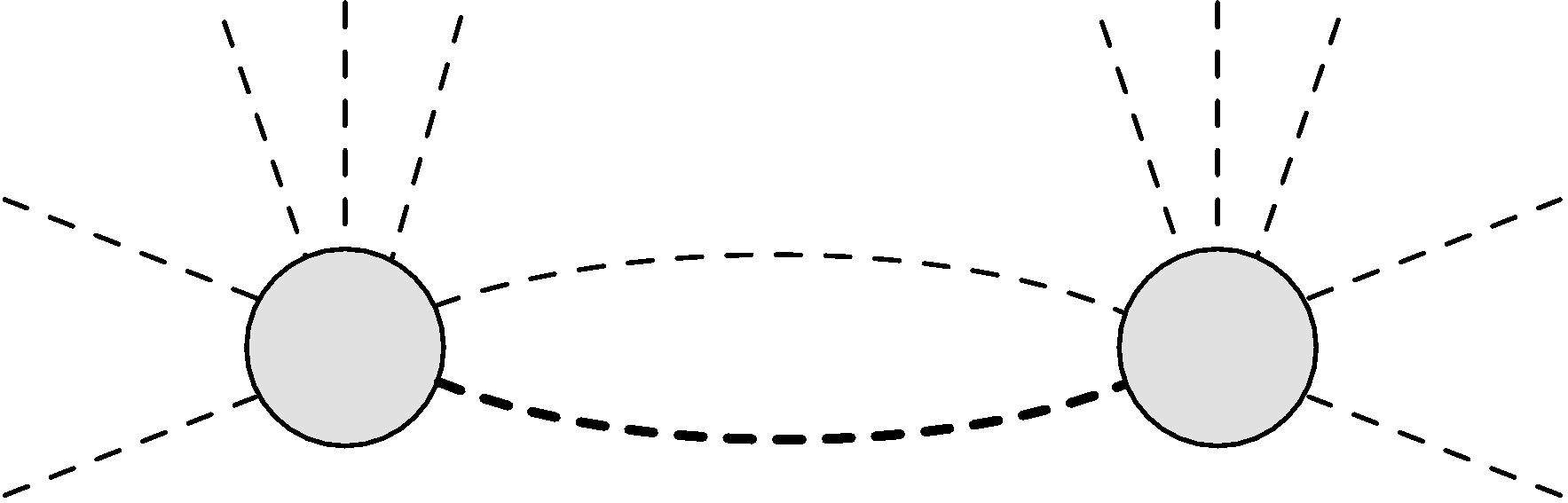}}
        \caption{Second order corrections to the relevant interactions with one loop closing between the two vertices. The grey ``blobs'' contain the first order vertex corrections. The double thick bosonic line is in the case when the two legs with momentum insertions connect among themselves.
 }
        \label{fig:Rel2OrdCancel}
\end{figure}

Let us begin by studying diagrams with only one loop closing between the two dressed vertices in which case we can demonstrate that the counting of the degree of divergence is the same using the two methods described above. 
The diagrams can be divided to three types: diagrams with 4 external fermion legs, diagrams with 2 external fermion legs and diagrams with no external fermion legs. All the  second order diagrams with one loop closing between the two vertices are depicted in figure  \ref{fig:Rel2OrdCancel}. The naive degree of divergence is logarithmic ($D=0$) for the diagrams in figures \ref{fig:Rel2OrdCancel-h}, \ref{fig:Rel2OrdCancel-i} and \ref{fig:Rel2OrdCancel-l} and convergent ($D<0$) for the rest. The blobs themselves are, after adding the first order quantum corrections, finite and equal to their tree level values. We therefore conclude that there are no quadratic divergences at second order with one loop closing between the two vertices.

Performing the analysis by the second method is also possible. For example taking the three divergent contributions of figures \ref{fig:Rel2OrdCancel-h}, \ref{fig:Rel2OrdCancel-i} and \ref{fig:Rel2OrdCancel-l} we obtain after explicitly performing the frequency integrals:
\begin{equation}\label{eq:2nd_order_rel_freq_int}
\begin{split}
&\frac{c_nc_m n! m!}{4}   \int \frac{d \omega_q d^2q}{(2\pi)^3}  
\\
& ~~~~~~~~ 
\left[
\frac{ g^2\, q^4}{\left({\omega_q}^2-(g\,q^2+m^2)^2 + i \epsilon\right) \left((\omega_q-\omega_k)^2 - (g \,(k-q)^2+m^2)^2 + i \epsilon \right)}\right.\\
& ~~~~~~~~ +  
\frac{ g^2 \, q^2(k-q)^2}{\left({\omega_q}^2-(g\, q^2+m^2)^2 + i \epsilon \right)  \left((\omega_q-\omega_k)^2 - (g\,(k-q)^2+m^2)^2 + i \epsilon \right)}\\
& ~~~~~~~~ - \left.\frac{1}{2} 
\frac{\omega_q(\omega_q-\omega_k)+(g\,q^2+m^2)(g\,(k-q)^2+m^2))}{\left({\omega_q}^2-(g\,q^2+m^2)^2 + i \epsilon \right) \left((\omega_q-\omega_k)^2 - (g\,(k-q)^2+m^2)^2 + i \epsilon \right)} \right]
\\
&
 = - \frac{i c_n c_m  n! m!}{8}  \int \frac{d^2 q}{(2\pi)^2}
\\
&
~~~~~~~~ 
\left[\frac{ g^2 q^2\left(q^2+(k-q)^2\right)\left(g\,q^2+g\,(k-q)^2+2m^2\right)}
{\left(g\,q^2+m^2\right)\left(g\,(k-q)^2+m^2\right)\left({\omega_k}^2-(g\,q^2+g\,(k-q)^2+2m^2)^2\right)} \right],
\end{split}
\end{equation}
where $\omega_k$, $k$ are the sums of external frequency and momenta entering one of the ``blobs'', and assuming there are $n-1$, $m-1$ external bosonic legs on the two vertices respectively. One can easily see that the last expression has a local logarithmic divergence which does not depend on $\omega_k$ or $k$, plus a finite part which depends on external momentum and frequency. The fact that the divergent contribution is local and does not, for example, contain inverse powers of the external frequency and momentum is important for the theory to be renormalizable. 

For higher number of loops closing between the two vertices we have not performed the frequency integrals explicitly, however we can perform the naive counting of the degree of divergence. The most diverging diagrams are 
similar to \ref{fig:Rel2OrdCancel-h}, \ref{fig:Rel2OrdCancel-i} and \ref{fig:Rel2OrdCancel-l}, but with additional (non thickened) bosonic lines attaching the two vertices. If we denote the number of loops by $l$, the naive degree of divergence is given by
\begin{equation}
D\leq -4(l+1)+4+4l = 0,
\end{equation}
which is again logarithmic.
It may not seem obvious at first why, when performing the frequency integrals and then imposing a momentum cutoff, diagrams which are logarithmically divergent by the weighted power counting will not contain e.g. contributions of the form $\Lambda^2/\omega_k$, where $\Lambda$ is the momentum cutoff and $\omega_k$ is the external frequency. Similar contributions arise when performing explicitly the frequency integral in \eqref{eq:2nd_order_rel_freq_int} and considering separately the contributions of each of the residues. For example, contributions which are proportional to $1/(\omega_k+g\,q^2-g\,(k-q)^2)$ and are less divergent than naively expected arise due to frequencies very close to poles whose location is not bounded in $q^2$. These contributions cancel when summing over the two residues. It has been argued by \cite{Anselmi:2007ri} that such contributions should not arise, and that in fact all needed counter terms are local. That is, when the subdivergences are removed, taking enough derivatives with respect to external momenta should make the integral overall convergent, hence, nonlocal divergences should not appear. 

\subsection{Quantum Corrections to the Marginal Interactions in 2+1 Dimensions}\label{subsec:Mar_quant}
\begin{figure}
   \centering
        \begin{subfigure}[b]{0.25\textwidth}
                \includegraphics[width=\textwidth]{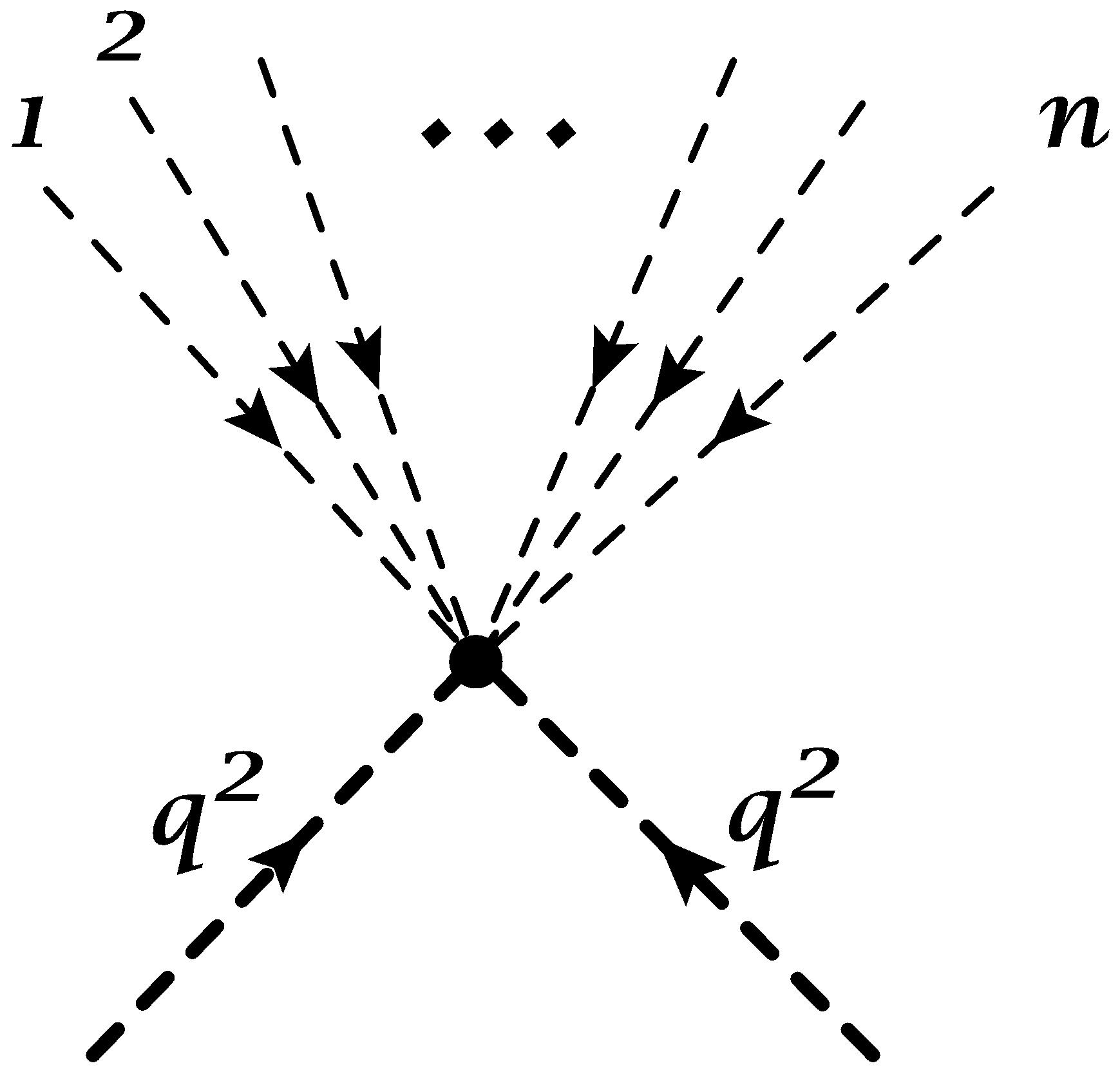}
                \caption{$\displaystyle -i\, \widetilde a_n g^2$}
                \label{fig:MargA}
        \end{subfigure}
~\quad        
        \begin{subfigure}[b]{0.25\textwidth}
                \includegraphics[width=\textwidth]{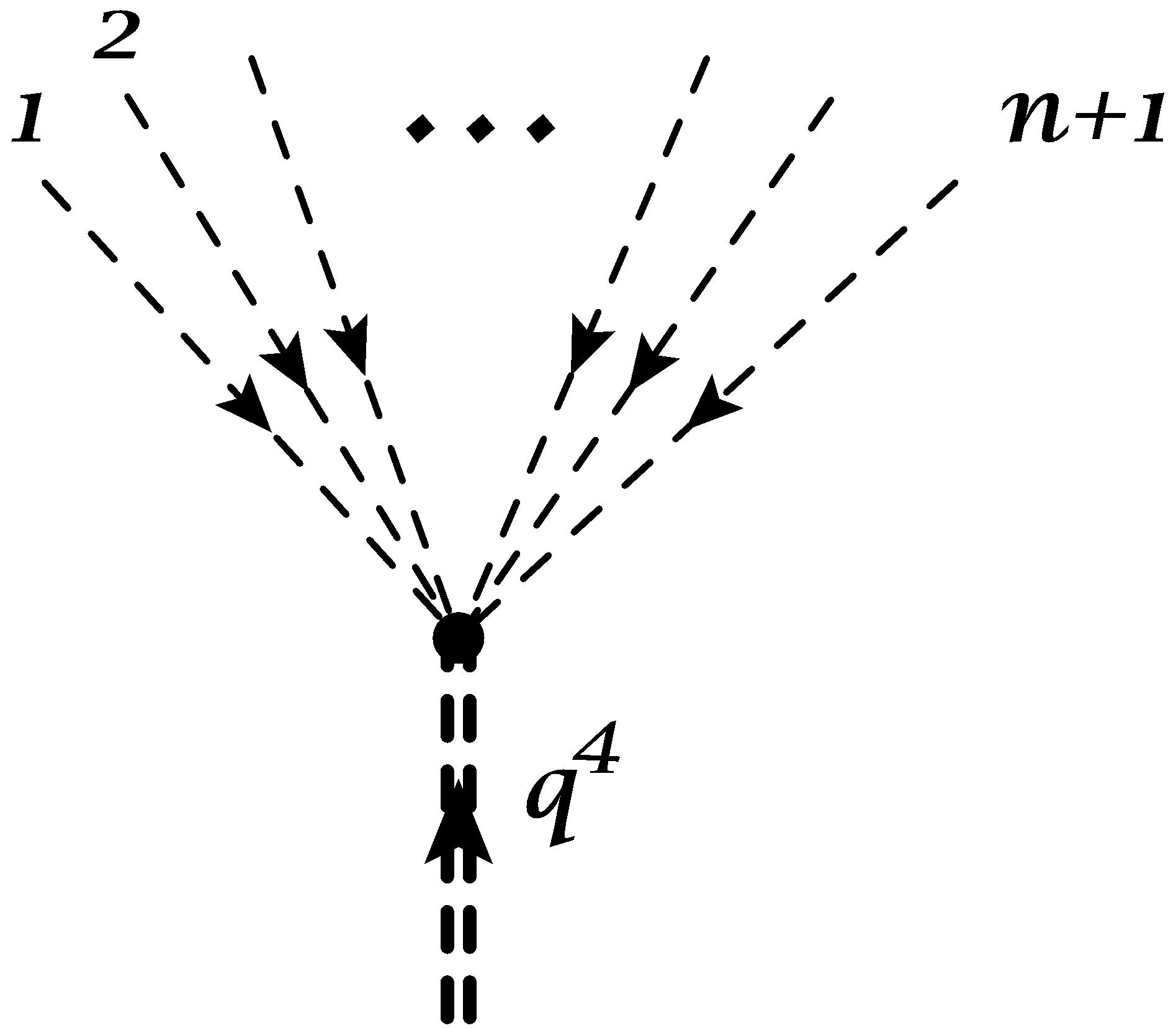}
                \caption{$\displaystyle -i\, \widetilde{a}_n\frac{1}{n+1}g^2$}
                \label{fig:MargB}
        \end{subfigure}
~\quad
        \begin{subfigure}[b]{0.25\textwidth}
                \includegraphics[width=\textwidth]{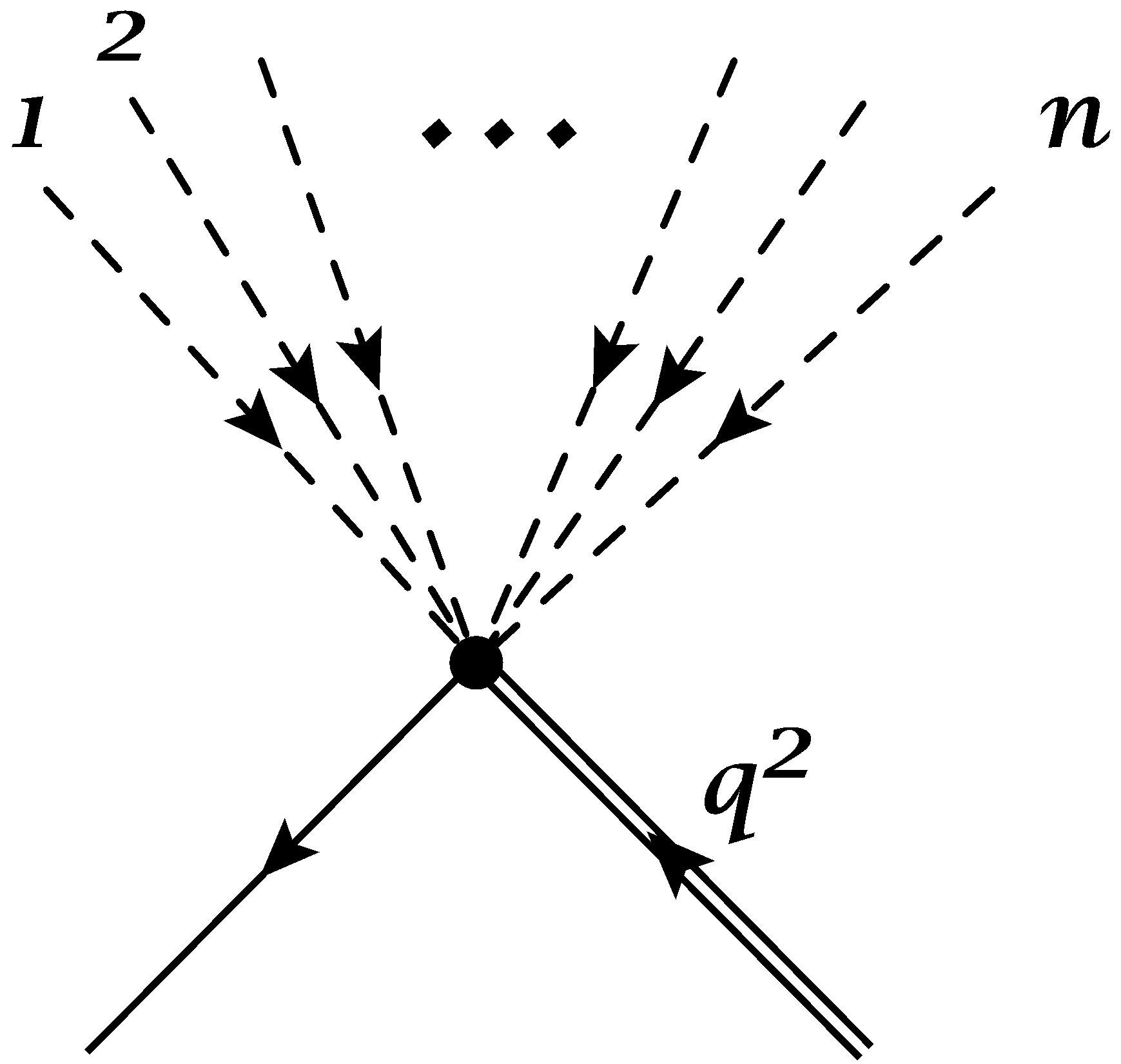}
                \caption{$\displaystyle - i \,\widetilde a_n g$}
                \label{fig:MargC}
        \end{subfigure}
\\
\vspace{1cm}
   \centering
        \begin{subfigure}[b]{0.25\textwidth}
                \includegraphics[width=\textwidth]{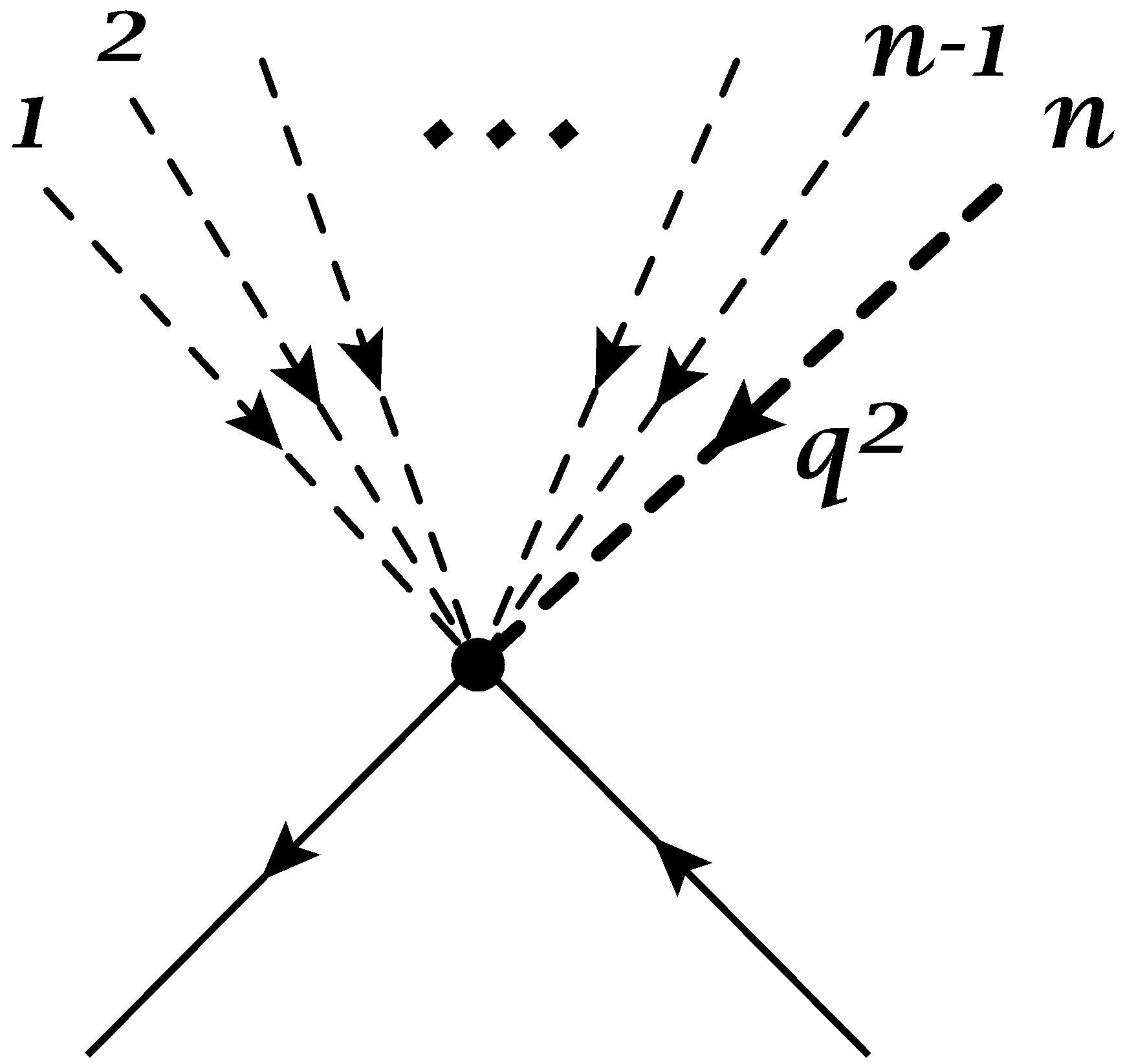}
                \caption{$\displaystyle - i \frac{n}{2} \,\widetilde a_n  g$}
                \label{fig:MargD}
        \end{subfigure}
~\quad        
        \begin{subfigure}[b]{0.25\textwidth}
                \includegraphics[width=\textwidth]{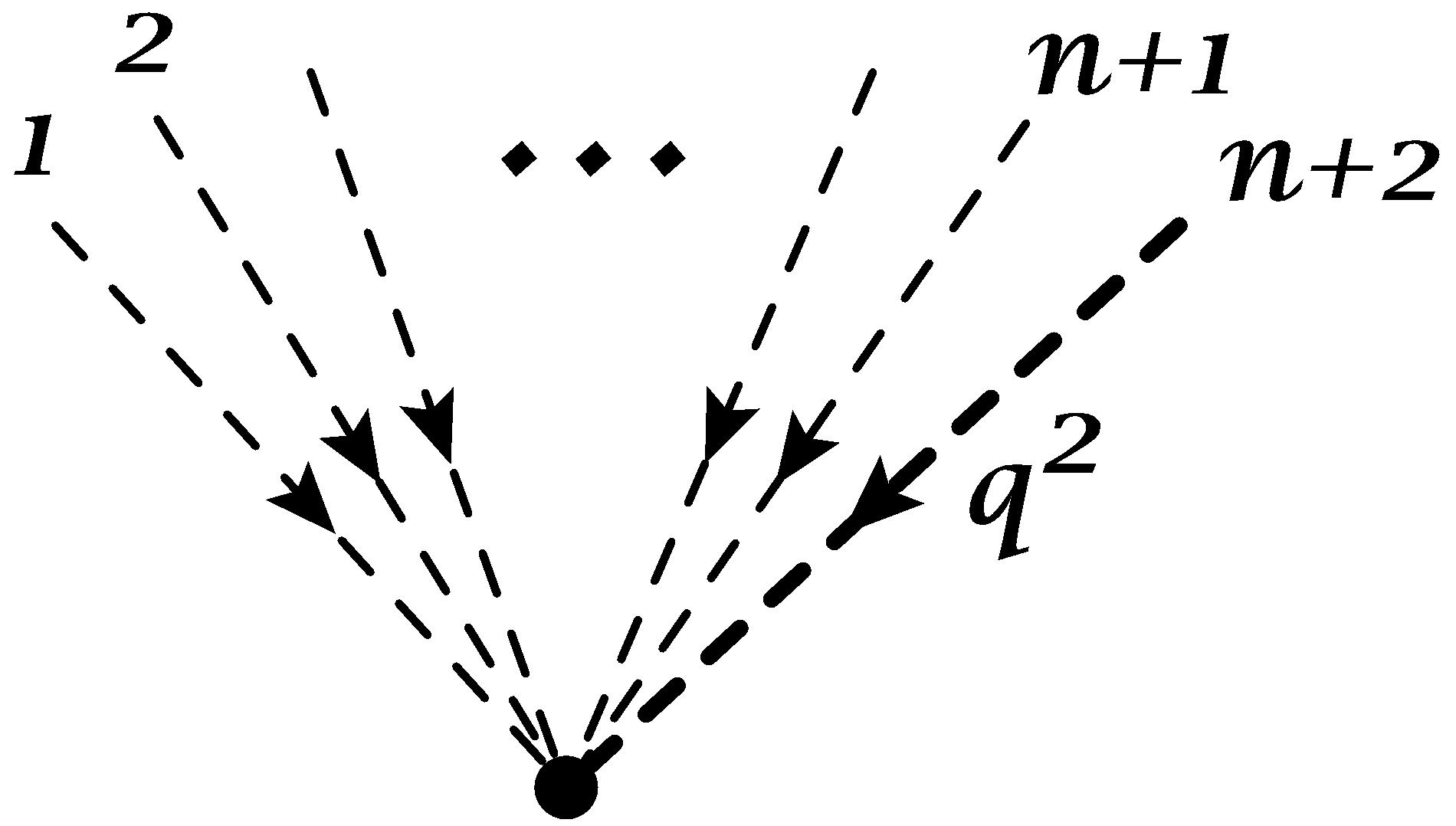}\\ \\
                \caption{$\displaystyle - i  \frac{n+2}{n+1} m^2 \,\widetilde a_n g $}
                \label{fig:MargE}
        \end{subfigure}
                       \caption{Feynman rules for the marginal interactions in renormalized perturbation theory. We kept only those contributions which are at most linear in the coupling constants. Momentum insertions are labeled on the vertex figures.}
        \label{fig:FeynMarg}    
\end{figure}

We now add the marginal interactions \eqref{eq:a2}. Since these interactions produce the relevant set of the previous subsection by quantum corrections we need to consider both at once. The solution for the auxiliary field $F$ introduce in this case also mixed interaction terms proportional to $c_1 a_n=2m^2a_n$ which contribute at first order in perturbation theory. We summarize here the additional interactions introduced by the marginal terms $a_n$ on top of those in \eqref{QC:RelL1Ord} at first order in perturbation theory:
\begin{equation}
\begin{split}
\mathcal L^{a+c} = \mathcal{L}^c & -   \sum\limits_{n = 1}^\infty  
 \widetilde a_n g \phi_r^{n - 1} \left[ g\left(\phi_r (\nabla^2\phi_r)^2+\frac{1}{n+1}\phi_r^2\nabla^4 \phi_r\right) -\frac{n+2}{n+1}m^2 \phi_r^2 \nabla^2\phi_r
\right]
\\& 
+  \sum\limits_{n = 1}^\infty  
 \widetilde a_n g \phi_r^{n - 1}  \left[\frac{n}{2} \nabla^2\phi_r \bar\psi_r\psi_r
 +\phi_r \bar\psi_r \nabla^2\psi_r
\right]
+ O\left(a_n^2,c_{n>1}^2,\delta^2\right),
 \end{split}
 \end{equation}
 where we have defined $\widetilde a_n \equiv a_n+\delta_{a_n} = a_n^{0}Z_\psi Z_\phi^{n/2}$. The relevant Feynman rules are in figure \ref{fig:FeynMarg}.

\subsubsection{Quantum Supersymmetry}
Let us define the standard quadratically divergent contribution:
\begin{equation}\label{QuantRel_BL_q2_full}
\mathcal{BL}_{q^2} \equiv \int \frac{ d\omega_q d^2q}{(2\pi)^3}
\, \frac{i}{2} \, \frac{q^2}{{\omega_q}^2-(g \, q^2+m^2)^2} ,
\end{equation}
and similarly:
\begin{equation}\label{QuantRel_BL_q4_full}
\mathcal{BL}_{q^4} \equiv \int \frac{ d\omega_q d^2q}{(2\pi)^3}
\, \frac{i}{2} \, \frac{q^4}{{\omega_q}^2-(g \, q^2+m^2)^2},
\end{equation}
in addition to the $\mathcal{BL}$ of equation \eqref{QuantRel_BL_full}.

We begin by studying the new contributions to the the fermion propagator \eqref{D_psi} with spatial external momentum $p$ from the vertices of figures \ref{fig:MargC} and \ref{fig:MargD}:
\begin{equation}
\delta \mathcal D_\psi = -i \left[g \sum\limits_k  \widetilde a_{2k}
\frac{(2k)!}{2^k k!}  (\mathcal{BL})^{k-1} \left((\mathcal{BL}) p^2 + k (\mathcal{BL}_{q^2}) \right) 
 \right].
\end{equation}
The corrections to the bosonic propagator \eqref{D_phi} from the vertices in figure \ref{fig:MargA}-\ref{fig:MargE} are given by:\footnote{Note, some of the diagrams in the sums proportional to $(k-1)$ begin to appear only at $k=2$.}
\begin{align}
\begin{split}
\delta \mathcal D_\phi^a = &\  -i g^2  \sum\limits_{k=1}^{\infty}  \widetilde a_{2k} \frac{(2k)!}{2^k k!}
(\mathcal{BL})^{k-2}  \left[2 p^4 (\mathcal{BL})^2 + 8k p^2 (\mathcal{BL}_{q^2}) (\mathcal{BL}) \right.
\\
&~~~~~~~~~~~~~~~~~~~~~~~~~~ \left.+ 2k  (\mathcal{BL}_{q^4}) (\mathcal{BL})
+4k(k-1) (\mathcal{BL}_{q^2})^2\right],
\end{split}
\\
\begin{split}
\delta \mathcal D_\phi^b = & \ -i g^2  \sum\limits_{k=1}^{\infty}  \widetilde a_{2k} \frac{(2k)!}{2^k k!}
(\mathcal{BL})^{k-1} \left[ 2p^4 (\mathcal{BL})
+ 2k (\mathcal{BL}_{q^4})
\right],
\end{split}
\\
\begin{split}
\delta \mathcal D_\phi^c = &\ 4 i g  \sum\limits_{k=1}^{\infty}  \widetilde a_{2k} \frac{(2k)!}{2^k (k-1)!}
(\mathcal{BL})^{k-1}  \left[g (\mathcal{BL}_{q^4}) +m^2 (\mathcal{BL}_{q^2}) \right],
\end{split}
\\
\begin{split}
\delta \mathcal D_\phi^d = &\ 4 i g  \sum\limits_{k=1}^{\infty}  \widetilde a_{2k} \frac{(2k)!}{2^k (k-1)!}
(\mathcal{BL})^{k-2}  \left[g (\mathcal{BL}_{q^2}) +m^2 (\mathcal{BL}) \right] \times \\
&~~~~~~~~~~~~~~~~~~~~~~~~~~~~~~~~~~~~~~~~~~
\left(p^2(\mathcal{BL})+(k-1)(\mathcal{BL}_{q^2})\right),
\end{split}
\\
\begin{split}
\delta \mathcal D_\phi^e = &\ -4 i g m^2  \sum\limits_{k=1}^{\infty}  \widetilde a_{2k} (k+1) \frac{(2k)!}{2^k k!}
(\mathcal{BL})^{k-1} \left[p^2 (\mathcal{BL}) + k (\mathcal{BL}_{q^2}) \right].
\end{split}
\end{align}
In total the bosonic propagator \eqref{D_phi} is corrected by:
\begin{equation}
\delta \mathcal{D}_\phi =  -4 i g \sum\limits_{k=1}^{\infty}  \widetilde a_{2k} \frac{(2k)!}{2^k k!}
(\mathcal{BL})^{k-1} (gp^2+m^2 )\left[p^2 (\mathcal{BL}) + k (\mathcal{BL}_{q^2}) \right].
\end{equation}
Note, that the highest divergent contributions $\mathcal{BL}_{q^4}$ and $(\mathcal{BL}_{q^2})^2$ completely cancel by the preserved 
supersymmetry.

Choosing:
\begin{equation}
\delta_g = - g \sum\limits_{k} \frac{(2k)!}{2^{k} k!} \, \widetilde a_{2k} \cdot  (\mathcal{BL})^k
\end{equation}
and:
\begin{equation}
\delta_m = - \sum\limits_{k} \frac{(2k+1)!}{2^{k+1} k!} \, \widetilde c_{2k+1} \cdot  (\mathcal{BL})^k - g \sum\limits_{k} \frac{(2k)!}{2^{k} (k-1)!} \, \widetilde a_{2k} \cdot  (\mathcal{BL})^{k-1} (\mathcal{BL}_{q^2}),
\end{equation}
yields again $\mathcal{D}_\phi=\mathcal{D}_\psi=0$. The fact that the same $\delta_m$ and $\delta_g$ allow the renormalization of various different terms in the Lagrangian is an indication that supersymmetry is preserved at the first perturbative order.

Similarly for the vertices \eqref{eq:fermion_blob} ($n\geq 2$):
\begin{equation}
\begin{split}
\delta \mathcal{M}_\psi^n
& \  = - \frac{i}{2} g \sum \limits_k \widetilde a_{n+2k-1} \frac{(n+2k-1)!}{2^{k-1} (k-1)!} (\mathcal{BL})^{k-1} ({\mathcal{BL}_{q^2}})
\\
&\ -\frac{i}{2} g 
\left( 2 p_{\psi_{_{in}}}^2 + \sum\limits_{ i \in bosonic\atop external\ legs} p_{\phi,i}^2 \right) \times 
\\
& ~~~~~~~~~~~~~~ \left[ \widetilde a_{n-1} (n-1)! + \sum \limits_k \widetilde a_{n+2k-1} \frac{(n+2k-1)!}{2^k k!} (\mathcal{BL})^k \right] 
\end{split}
\end{equation}
and \eqref{eq:scalar_blob}:
\begin{equation}
\begin{split}
&\delta\mathcal{M}_\phi^n =  - ig  \left(m^2 (n+1)+ \sum\limits_{i \in bosonic\atop external\ legs} g p_{\phi,i}^2 \right) \times
\\ & ~~~~~~~~~~~~~~~~~~~~
\left[ \sum \limits_{k=1}^{\infty}
\, \widetilde a_{n+2k-1}  \frac{(n+2k-1)!}{2^{k-1} (k-1)!} (\mathcal{BL})^{k-1} ({\mathcal{BL}_{q^2}}) 
\right]
\\
&\ - i g 
\left( g \sum\limits_{ i \in bosonic\atop external\ legs} p_{\phi,i}^4  + 2 g \sum\limits_{ i <j \in bosonic\atop external\ legs} p_{\phi,i}^2 \,  p_{\phi,j}^2
+ m^2(n+1) \sum\limits_{ i \in bosonic\atop external\ legs} p_{\phi,i}^2  \right)
\times
\\&
~~~~~~~~~~~~~~~~~~~~ \left[ \widetilde a_{n-1} (n-1)! + \sum \limits_k \widetilde a_{n+2k-1} \frac{(n+2k-1)!}{2^k k!} (\mathcal{BL})^k \right] .
\end{split}
\end{equation}
Choosing:
\begin{equation}
\delta_{a_n} = -\frac{1}{(n-1)!} \sum \limits_k \widetilde a_{n+2k-1} \frac{(n+2k-1)!}{2^k k!} (\mathcal{BL})^k 
\end{equation}
and
\begin{equation}
\delta_{c_n}  = - \frac{1}{n!}\sum \limits_{k=1}^{\infty}
\, (\mathcal{BL})^{k-1}  \frac{(n+2k)!}{2^k k!} \left[ \widetilde c_{n+2k}  (\mathcal{BL})
+
 g \widetilde a_{n+2k-1} \frac{2k}{n+2k} ({\mathcal{BL}_{q^2}})
\right]
\end{equation}
we are able to renormalize all vertices, preserving supersymmetry. 
We can also see that the relevant interactions do not correct the marginal coupling, justifying in retrospect our separate study of the relevant interactions in subsection \ref{subsec:Rel_quant}.

\subsection{Lifshitz Supersymmetry in 3+1 Dimensions}\label{app:3+1susyLifshitz} 
In 3+1 dimensions with $z=2$ we find for the interactions \eqref{4dactionGen} that again  supersymmetry is preserved at first order in perturbation theory. We also deduce that scale invariance is broken at the quantum mechanical level at the second order in perturbation theory, containing two vertices and two loops.
\subsubsection{Quantum Supersymmetry}\label{subsec:Conserv_4d_susy}

The definitions of the counter terms remain as in \eqref{eq:RedefRelParams} and the Feynman rules remain those of figure \ref{fig:FynRelev2} with all $c_n$ with $n>5$ set to zero.
The renormalization procedure is very similar to the $2+1$ dimensional case. The only changes are the cutoff of the (formerly) infinite interaction series and the fact that the  integration measure becomes $d\omega_q d^3 q/(2\pi)^4$. The standard divergent integral $\mathcal{BL}$ of equation \eqref{QuantRel_BL_full} is replaced with:
\begin{equation}
\label{bl4d}
\mathcal{BL}^{(4d)} \equiv \int \frac{ d\omega_q d^3q}{(2\pi)^3} {\mkern 1mu} \frac{i}{2} {\mkern 1mu} \frac{1}{{\omega_q}^2 - (g{q^2} + m^2)^2}= \frac{\Lambda }{4\pi g} + {\rm{finite}}.
\end{equation}

The first order quantum corrections are still captured by \eqref{D_phi}--\eqref{c_n_corr} with the replacement ${\cal B}{\cal L} \to {\cal B}{{\cal L}^{(4d)}} $ and $c_n$ set to zero for all $n>5$ (only $k=1,2$ in the sums). 

We detail here explicitly the quantum corrections to demonstrate what happens when the interaction series is cutoff.
The quantum corrections to the fermionic and bosonic propagators are given by:
\begin{equation}
\mathcal{D}_\phi  =  - i(g p^2 + m^2) \left[ 4 (\delta _m+p^2\delta_g) + 
6 \, \mathcal{BL}^{(4d)} \left(\tilde c_3 +5 \tilde c_5 \mathcal{BL}^{(4d)} \right) \right]
\end{equation}
\begin{equation}
\mathcal{D}_\psi  =  - i\left[ (\delta_m+p^2\delta_g) +  \frac{6}{4} \mathcal{BL}^{(4d)} \left(\tilde c_3 +5 \tilde c_5 \mathcal{BL}^{(4d)} \right) \right]
\end{equation}
Where $p$ is the spatial external momentum. Therefore setting 
\begin{equation}\label{delta_m_3d}
{\delta _m} =  - \frac{6}{4} \mathcal{BL}^{(4d)} \left(\tilde c_3 +5 \tilde c_5 \mathcal{BL}^{(4d)} \right), \qquad \delta_g=0
\end{equation}
yields $D_\phi=D_\psi=0$ for all $p^2$ (which is consistent with the conditions on the propagator's poles and residues). We have again defined $\widetilde c_n = c_n+\delta_{c_n}$. It is not clear a priori why $\delta_g$ does not get corrected by the marginal coupling $c_5$. The $\delta_{c_n}$ will be fixed by the quantum corrections to the vertices. 
The contributions to a scattering amplitudes ${\cal M}_\psi^n$ with 2 external fermionic legs and $n-1$ external bosonic legs ($n\geq 2$) are  given by:
\begin{equation}
\begin{array}{l}
{\cal M}_\psi ^2 =  - i\left(\tilde c_2 + 6 \tilde c_4 \, \mathcal{BL}^{(4d)}\right),\quad
{\cal M}_\psi ^3 =  - 3i \left( \tilde c_3 + 10 \tilde c_5 \, \mathcal{BL}^{(4d)} \right),
\\ 
{\cal M}_\psi ^4 =  - 12 i \tilde c_4, 
~~\qquad \qquad \qquad 
{\cal M}_\psi ^5 =  - 60 i \tilde c_5.
\end{array}
\end{equation} 
The scattering amplitudes with $n+1$ external bosonic legs ${\cal M}_\phi^n$ equal:
 \begin{equation}
 \begin{array}{l}
{\cal M}_\phi ^2 =  - 2i\left( {3{m^2} + g\sum\limits_{i = 1}^3 {p_i^2} } \right)
\left(\tilde c_2 + 6 \tilde c_4 \, \mathcal{BL}^{(4d)}\right),
\\
{\cal M}_\phi ^3 =  - 6i\left( {4{m^2} + g\sum\limits_{i = 1}^4 {p_i^2} } \right)
\left( \tilde c_3 + 10 \tilde c_5 \, \mathcal{BL}^{(4d)} \right),
\\
{\cal M}_\phi ^4 =  - 24 i\left( {5{m^2} + g\sum\limits_{i = 1}^5 {p_i^2} } \right)\tilde c_4,\qquad
{\cal M}_\phi ^5 =  - 120 i\left( {6{m^2} + g\sum\limits_{i = 1}^6 {p_i^2} } \right)\tilde c_5.
\end{array}
 \end{equation}
To make an appropriate choice of the corrections one needs to properly define the coupling constants. Choosing e.g. that the appropriate amplitude equals its tree level value for some given values of the external momenta we get:
 \begin{equation}\label{delta_c_3d}
 \begin{array}{l}
{\delta _{{c_2}}} =  - 6 c_4 {\cal B}{{\cal L}^{\left( {4d} \right)}}, \qquad 
{\delta _{{c_3}}} =  - 10 c_5 {\cal B}{{\cal L}^{\left( {4d} \right)}}, \qquad
{\delta _{{c_4}}} = {\delta _{{c_5}}} = 0,
\end{array}
 \end{equation}
which can be plugged back into \eqref{delta_m_3d} to obtain:
\begin{equation}\label{delta_m_3d__2}
{\delta _m} =  - \frac{6}{4} \mathcal{BL}^{(4d)} \left(c_3 -5 c_5 \mathcal{BL}^{(4d)} \right),
\end{equation}
which is both quadratically and linearly divergent. This choice of corrections will keep the structure of the original vertices for any value of the external momentum at the first perturbative order. This is due to the fact that the loop integrals do not depend on the external momentum at this order. Note again that the fact that the independent corrections can all be reabsorbed in the same $\delta_m$ and $\delta_{c_n}$ indicates that supersymmetry is preserved at this order.

 \subsubsection{Breakdown of Lifshitz Scale Invariance}\label{Breaking_Scale_4d_3d}
Using the the stochastic quantization approach of section \ref{subsec:detbal} we can study the breakdown of scale invariance in the $z=2$, $3+1$ dimensional supersymmetric theory from the corresponding (classically scale invariant) three-dimensional relativistic bosonic theory with only $c_5$ being non-zero. The three- dimensional relativistic theory was previously studied in \cite{McKeon:1992cs,Pisarski:1982vz}.

\begin{figure}[ht!]
\centering
\includegraphics[width=60mm]{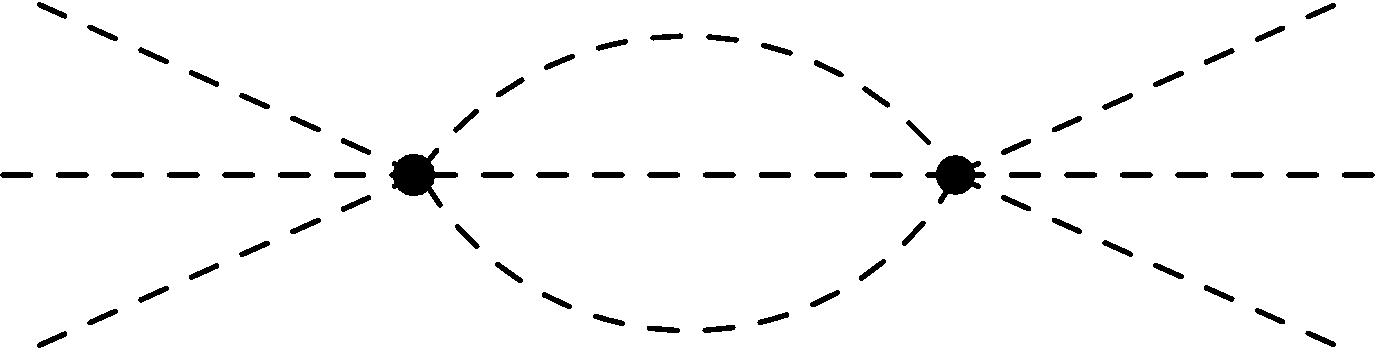}
\caption{The first non-vanishing contribution to the $\beta$ function of the coupling constant $c_5$. \label{fig:beta_3d_4d}}
\end{figure}

We have already seen that $c_5$ does not run at the first perturbative order. However, at the second perturbative order $c_5$ has a non vanishing beta function and hence scale invariance is broken. The contribution to the beta function comes from the diagram depicted in figure \ref{fig:beta_3d_4d} and reads:
\begin{equation}
\label{beta_func_3d}
\beta \left( {{c_5}} \right) \equiv \mu \frac{{\partial {c_5}}}{{\partial \mu }} = \left( \frac{5 c_5}{2\pi } \right)^2 > 0,
\end{equation}
This diagram is logarithmically divergent and has a logarithmic dependence on the external momentum. Therefore it creates a non-trivial $\beta$ function, which leads to a breaking of scale invariance. 

\subsection{A Comparison with Stochastic Quantization}
\label{subsec:Comp_Detailed_Balance}
One can use the stochastic quantization framework and compare our results for $z=2$ with those of the corresponding relativistic bosonic theory \eqref{DetailedBalance1}.
In this subsection we compare some of the properties of the two theories. In particular we show that the superficial degree of divergence matches.

In the $2+1$, $z=2$ supersymmetric model with only relevant couplings $c_n$ the highest degree of divergence obtained was logarithmic (see subsection \ref{subsec:Rel_quant}). This is also the case for the corresponding two dimensional relativistic case of equation \eqref{DetailedBalance1} with only $c_n \neq 0$ whose naive degree of divergence is given by:
\begin{equation}
D = 2(1-V) , 
\end{equation}
where $V$ is the number of vertices. This is at most logarithmically divergent.

In the case with marginal couplings $a_n$ the degree of divergence was at most quadratic in the $z=2$ Lifshitz supersymmetric case of subsection \ref{subsec:Mar_quant}. In the corresponding relativistic theory of section \ref{subsec:detbal} with nonzero $a_n$ we obtain the same behavior since each vertex can be accompanied by two powers of momentum thus correcting the superficial degree of divergence to be at most 2.
The fact that the relevant set of coupling constants does not generate corrections to the marginal set also matches the relativistic behavior,
where it can be explained by dimensional arguments.

The same analysis can be performed for the $3+1$ dimensional, $z=2$ supersymmetric model by studying the quantum corrections to the three-dimensional bosonic theory. 
Here the superficial degree of divergence is given by:
\begin{equation}
D = 3- \frac{N}{2} - \frac{3}{2} V_{c_2} - V_{c_3} -\frac{1}{2} V_{c_4},
\end{equation}
where $N$ is the number of external legs, $V_{c_i}$ is the number of vertices of type $c_i$. 
One can compare the corrections in equations 
\eqref{delta_c_3d} - \eqref{delta_m_3d__2} obtained in the presence of any of the vertices. The naive counting matches exactly.

\section{Summary and Outlook}\label{sec:Sum}

In this work we considered classical and quantum aspects of Lifshitz supersymmetry. In the cases that we studied we found that quantum corrections
preserve supersymmetry, while there are indications for a breakdown of the scale symmetry, at least in $3+1$ dimensions.
The models that we considered are intimately related to those that appear in the framework of stochastic quantisation with a detailed
balance condition,  making them natural in the presence of a stochastic noisy background. 

In the work we encountered interesting aspects of Lifshitz supersymmetry that deserve further studies.
When making an attempt to formulate Lifshitz supersymmetry via superfields one encounters a difficulty. A natural generalization of the standard supersymmetry differential operator $Q$ for Lifshitz field theories with $z\neq 1$ contains more then one space derivative. Such an operator does not satisfy a standard product rule (e.g. $[Q,\Phi^2] \neq 2 \Phi [Q,\Phi]$). This makes it challenging to construct the Lagrangians and interaction terms via a superspace formalism. 

The condition for supersymmetry breaking in Lifshitz theories is similar to that of the relativistic case
$E \equiv \left\langle 0 \right| H \left| 0 \right\rangle > 0$.
The study of spontaneous Lifshitz supersymmetry breaking and the structure of the goldstinos is an important direction to pursue.
Going beyond the leading orders in perturbation theory in the study of quantum Lifshitz symmetry is also of interest.
In the models that we considered it is likely that the stochastic quantisation framework will prove useful.

A complete classification of supersymmetric Lifshitz theories with general representations of the fermions under spatial rotations
is clearly a desirable objective.
Lastly, it will be of much interest to analyse the potential consequences of Lifshitz supersymmery if manifested in condensed matter
systems.

\section*{Acknowledgement}
We would like to thank Cobi Sonnenschein, Shimon Yankielowicz, Guy Gur Ari, Itamar Shamir, and especially
 Igal Arav  and Lorenzo Di Pietro for valuable discussions and comments.  We also thank Adam Chapman for suggesting a combinatorial identity that simplified the results of section  \S \ref{sec:FreeLifshitzSUSY}.
This work is supported in part by the I-CORE program of Planning and Budgeting Committee (grant number 1937/12), the US-Israel Binational Science Foundation, GIF and the ISF Center of Excellence.

\appendix

\section{Notations and Conventions for Two Component Real Fermions}
\label{app:conventions}
We follow the conventions of \cite{RuizRuiz:1996mm}. 
We use the following (real) representation of the Gamma matrices: 
\begin{align}
\begin{split}
&(\gamma^\mu)^\alpha{}_\beta = (-i\sigma^2,\sigma^1,\sigma^3),
\\
&(\gamma^\mu)_{\alpha\beta} = (-I,-\sigma^3,\sigma^1), 
\\
&(\gamma^\mu)_\alpha{}^\beta = (i\sigma^2,\sigma^1,\sigma^3),
\\
&(\gamma^\mu)^{\alpha\beta} = (-I,\sigma^3,-\sigma^1).
\end{split}
\end{align}
The sigma matrices are:
\begin{equation}
\sigma^1 = \left( \begin{matrix}
0 & 1\\1 & 0  
\end{matrix}\right),
\quad
\sigma^2 = \left( \begin{matrix}
0 & -i\\i & 0  
\end{matrix}\right),
\quad
\sigma^3 = \left( \begin{matrix}
1 & 0\\0 & -1  
\end{matrix}\right).
\end{equation}
Note that:
\begin{equation}
\left\{ {{\gamma ^\mu },{\gamma ^\nu }} \right\} = 2{\eta ^{\mu \nu }},
\end{equation}
where ${\eta ^{\mu \nu }} = diag\left( { - 1,1,1} \right)$. 

The spinors are raised and lowered using the \emph{northwest-southeast} convention:
\begin{equation}
\psi ^\alpha = \epsilon ^{\alpha \beta }\psi _\beta,
\qquad
\psi_\alpha  = \psi ^\beta  \epsilon _{\beta \alpha },
\end{equation}
where $\epsilon_{12} = \epsilon^{12}=1$. 
This implies $V^\alpha U_\alpha = - V_\alpha U^\alpha$ for any two spinors $U$ and $V$. 
The order of contractions is \emph{lower-upper} unless specified otherwise. 
Transposition changes the height of an index.

We define:
\begin{equation}
\bar \psi  = {\psi ^T}{\sigma ^2}.
\end{equation}
In terms of explicit spinor components:
\begin{equation}\label{app:compts}
\psi_\alpha= \left(\begin{matrix} \psi_1 \\ \psi_2 \end{matrix} \right)
,~~~~~~
\psi^\alpha=\epsilon^{\alpha\beta} \psi_\beta =  \left(\begin{matrix} \psi_2 \\ -\psi_1 \end{matrix} \right),
\end{equation}
\begin{equation*}
(\bar\psi)^\alpha = (\psi_\alpha)^T \sigma^2 = 
\left(
\begin{matrix} \psi_1 & \psi_2 
\end{matrix} \right)
 \left( \begin{matrix}
0 & -i\\i & 0  
\end{matrix}\right)
= i
\left(
\begin{matrix} \psi_2 & -\psi_1 
\end{matrix} \right) \cong i\psi^\alpha,
\end{equation*}
therefore
\begin{equation*}
\bar \psi_\alpha = \bar\psi^\beta \epsilon_{\beta\alpha} = i \left(\begin{matrix} \psi_1 \\ \psi_2 \end{matrix} \right) = i \psi_\alpha,
\end{equation*}
where $\cong$ stands for component by component equality. This dependence makes sense as we are dealing with real spinors i.e. $\psi_1^* = \psi_1$ and $\psi_2^* = \psi_2$. We work under the following convention $(\eta\chi)^*=\chi^* \eta^*$ for the complex conjugation of fermions.

We will use the following identities for real spinors:
\begin{equation}\label{Sp_ID}
\begin{split}
\bar \psi \chi & = \bar \chi \psi,
\\
\bar \psi {\gamma ^\mu }\chi & =  - \bar \chi {\gamma ^\mu }\psi,
\\
(\bar \psi \eta) \chi_\alpha & = -(\bar \chi \eta) \psi_\alpha - (\bar \chi \psi) \eta_\alpha,
\end{split}
\end{equation}
which can be easily checked by the explicit component notation.


\begin{thebibliography}{999}
 \bibitem{Sachdev:2011cup} 
  S.~Sachdev,
  ``Quantum Phase Transitions'',
Cambridge University Press (2011).
   
\bibitem{Martin:1997ns}
  S.~P.~Martin,
  ``A Supersymmetry primer'',
  Adv.\ Ser.\ Direct.\ High Energy Phys.\  {\bf 21} (2010) 1
  [\href{http://arxiv.org/abs/hep-ph/9709356}{hep-ph/9709356}].



\bibitem{Parisi:1982ud} 
  G.~Parisi and N.~Sourlas,
  ``Supersymmetric Field Theories and Stochastic Differential Equations'',
  Nucl.\ Phys.\ B {\bf 206}, 321 (1982),
  [\href{http://www.sciencedirect.com/science/article/pii/0550321382905387}{NPB.206/321}].

\bibitem{Dijkgraaf:2009gr} 
  R.~Dijkgraaf, D.~Orlando and S.~Reffert,
  ``Relating Field Theories via Stochastic Quantization'',
  Nucl.\ Phys.\ B {\bf 824}, 365 (2010),
  [\href{http://arxiv.org/abs/0903.0732}{hep-th/0903.0732}].

\bibitem{Horava:2008ih} 
  P.~Horava,
  ``Membranes at Quantum Criticality'',
  JHEP {\bf 0903}, 020 (2009),
  [\href{http://arxiv.org/abs/0812.4287}{hep-th/0812.4287}]. 

 \bibitem{Orlando:2009az}
  D.~Orlando and S.~Reffert,
  ``On the Perturbative Expansion around a Lifshitz Point'',
  Phys.\ Lett.\ B {\bf 683} (2010) 62
  [\href{http://arxiv.org/abs/0908.4429}{hep-th/0908.4429}].

  
\bibitem{Xue:2010ih}
  W.~Xue,
  ``Non-relativistic Supersymmetry'',
  [\href{http://arxiv.org/abs/1008.5102}{hep-th/1008.5102}].
 
\bibitem{Gomes:2014tua}
  M.~Gomes, J.~R.~Nascimento, A.~Y.~Petrov and A.~J.~da Silva,
  ``On the Horava-Lifshitz-like extensions of supersymmetric theories'',
   [\href{http://arxiv.org/abs/1408.6499}{hep-th/1408.6499}].

\bibitem{Damgaard:1987rr} 
  P.~H.~Damgaard and H.~Huffel,
  ``Stochastic Quantization'',
  Phys.\ Rept.\  {\bf 152}, 227 (1987).
[\href{http://homepage.univie.ac.at/helmuth.hueffel/PhysRep.pdf}{PhysRep:152.227.1987}] 

\bibitem{Anselmi:2007ri} 
  D.~Anselmi and M.~Halat,
  ``Renormalization of Lorentz violating theories'',
  Phys.\ Rev.\ D {\bf 76}, 125011 (2007)
  [\href{http://arxiv.org/abs/0707.2480}{hep-th:0707.2480}]. 


\bibitem{Fitzpatrick:2012ww}
  A.~L.~Fitzpatrick, S.~Kachru, J.~Kaplan, E.~Katz and J.~G.~Wacker, ``A New Theory of Anyons'',
  [\href{http://arxiv.org/abs/1205.6816}{hep-th/1205.6816}].

\bibitem{Pisarski:1982vz} 
  R.~D.~Pisarski,
  ``Fixed point structure of $\phi^6$ in three-dimensions at large N'',
  Phys.\ Rev.\ Lett.\  {\bf 48}, 574 (1982),
  [\href{http://journals.aps.org/prl/abstract/10.1103/PhysRevLett.48.574}{PhysRevLett.48.574}].

\bibitem{McKeon:1992cs} 
  D.~G.~C.~McKeon and G.~Tsoupros,
  ``Perturbative evaluation of renormalization group functions in massive three-dimensional $\phi^6$ theory'',
  Phys.\ Rev.\ D {\bf 46}, 1794 (1992),
  [\href{http://journals.aps.org/prd/abstract/10.1103/PhysRevD.46.1794}{PhysRevD.46.1794}].


\bibitem{RuizRuiz:1996mm}
  F.~Ruiz Ruiz and P.~van Nieuwenhuizen,
  ``Lectures on supersymmetry and supergravity in 2+1 dimensions and regularization of supersymmetric gauge theories'',
 [\href{http://lpsc.in2p3.fr/d0/members/besson/documents/SUSY/GENERAL/susy_ruiz.ps}{Ruiz.ps}].


   
\end{thebibliography}
\end{document}